\documentclass[11pt,a4paper]{article}
\pdfoutput=1

\usepackage{jheppub}
\usepackage{latexsym}
\usepackage{revsymb}
\usepackage{multirow}
\usepackage{color}
\usepackage[usenames,dvipsnames,svgnames,table]{xcolor}
\usepackage{lipsum}
\usepackage{footnote}
\usepackage{graphicx}
\usepackage{epsfig}  
\usepackage{epsf}    
\usepackage{dcolumn}
\usepackage{bm}
\usepackage{dcolumn}
\usepackage{textcomp}
\usepackage{float}
\usepackage{subfig}
\usepackage{lineno}
\usepackage{hypcap}
\usepackage[]{hyperref}

\hypersetup{
  bookmarks=true,         
  unicode=false,          
  pdftoolbar=true,        
  pdfmenubar=true,        
  pdffitwindow=true,     
  pdfstartview={FitH},    
  pdfsubject={Neutrino Oscillations Phenomenology},   
  pdfnewwindow=true,      
  pdfcreator={RevTeX},
  colorlinks=true,       
  linkcolor=red,          
  citecolor=blue,        
  filecolor=black,      
  urlcolor=blue,           
}

\newcommand{\eg}{{\it e.g.}}
\newcommand{\ie}{{\it i.e.}}
\newcommand{\etc}{{\it etc.}}

\newcommand{\beq}{\begin{equation}}
\newcommand{\eeq}{\end{equation}}
\newcommand{\beqa}{\begin{eqnarray}}
\newcommand{\eeqa}{\end{eqnarray}}
\newcommand{\tij}{{\theta_{ij}}}
\newcommand{\tx}{{\theta_{12}}}
\newcommand{\ty}{{\theta_{13}}}
\newcommand{\tz}{{\theta_{23}}}
\newcommand{\ta}{{\theta_{14}}}
\newcommand{\tb}{{\theta_{24}}}
\newcommand{\tc}{{\theta_{34}}}

\newcommand{\dcp}{\delta_{\mathrm{CP}}}

\newcommand{\pab}{P(\alpha \rightarrow \beta)}

\newcommand{\dxx}{\Delta\chi^2}

\newcommand{\da}{\delta_{13}}
\newcommand{\db}{\delta_{24}}
\newcommand{\dc}{\delta_{34}}

\preprint{FERMILAB-PUB-17-276-T}

\title{What measurements of neutrino neutral current events can reveal}

\author[a]{Raj Gandhi,}
\author[b]{Boris Kayser,} 
{\author[c]{Suprabh Prakash,}
\author[a]{Samiran Roy$\,$} 

\affiliation[a]{Harish-Chandra Research Institute, HBNI, Chhatnag Road, Jhunsi,
Allahabad 211019, India}
\affiliation[b]{Theoretical Physics Department, Fermilab, P.O. Box 500, Batavia,
IL 60510 USA}
\affiliation[c]{Instituto de F\'isica Gleb Wataghin - UNICAMP, 13083-859, Campinas, S\~ao Paulo, Brazil}

\emailAdd{raj@hri.res.in}
\emailAdd{boris@fnal.gov}
\emailAdd{samiranroy@hri.res.in}
\emailAdd{sprakash@ifi.unicamp.br}

\begin{abstract}
{We show that neutral current (NC) measurements at neutrino detectors can play a valuable role in the search for new physics. Such measurements have certain intrinsic features and advantages  that can fruitfully be combined with the usual  well-studied charged lepton detection channels in order to probe the presence of new interactions or new light states. In addition to the fact that NC events are immune to uncertainties in standard model neutrino mixing and mass parameters, they can have small matter effects  and superior rates since all three flavours participate. We also show, as a general feature, that  NC measurements provide access to different combinations of CP phases and mixing parameters compared to CC measurements at both long and short baseline experiments.  Using the Deep Underground Neutrino Experiment (DUNE) as an illustrative setting,  we demonstrate the capability of NC measurements to break degeneracies arising in CC measurements, allowing us, in principle, to distinguish between new physics that violates three flavour unitarity  and that which does not. Finally, we show that NC measurements can enable us to restrict new physics parameters that are not easily constrained by CC measurements.

}
\end{abstract}

\keywords{}
\arxivnumber{}

\begin{document}
\maketitle
\flushbottom

\section{Introduction}

The major goals of present-day and near-future neutrino oscillation
experiments are: a) the determination of the neutrino mass hierarchy (MH)  and 
b) the discovery and possible measurement of the magnitude of CP violation (CPV)
in the lepton sector. In addition,  ancillary goals include
 making  increasingly precise determinations of neutrino mass-squared
differences,  $\delta m^{2}_{ij} = m^{2}_{i} - m^{2}_{j}$ (i, j = 1, 2, 3 \& $i
\neq j$) and mixing angles $\tij$. Recent status reviews may be found in 
\cite{Cao:2017hno,TalkbyAlexSousa,TalkbyWalterWinter}.

The  capability for  increased 
precision in neutrino experiments has recently  led to the formulation of another 
important line of inquiry: the search for new physics at neutrino detectors, and its 
identification and disentanglement from physics related to the standard model with 
three generations of massive neutrinos. Examples of recent work in this direction may be found in 
\cite{
Esmaili:2013cja,
Chatterjee:2014gxa,
Klop:2014ima,
Choubey:2015xha,
Blennow:2015nxa,
Berryman:2015nua,
Parke:2015goa,
Gandhi:2015xza,
deGouvea:2015ndi,
Coloma:2015kiu,
Forero:2016cmb,
Liao:2016reh,
Berryman:2016szd,
Masud:2016bvp,
Agarwalla:2016xxa,
Choubey:2016fpi,
Miranda:2016wdr,
Coloma:2016gei,
Ge:2016xya,
Agarwalla:2016xlg,
Masud:2016gcl,
Blennow:2016etl,
Agarwalla:2016fkh,
Dutta:2016glq,
Verma:2016nfi,
Dutta:2016czj,
Blennow:2016jkn,
Dutta:2016eks,
Escrihuela:2016ube,
Deepthi:2016erc,
Cianci:2017okw,
Rout:2017udo,
Choubey:2017cba,
Masud:2017bcf, 
Ghosh:2017atj,
Choubey:2017dyu,
Coloma:2017zpg,
Coloma:2017ptb}.
It is the purpose of this work to bring out facets of neutral current (NC) measurements at neutrino detectors that can aid in furthering efforts in this direction either on their own or when employed in synergy with other measurements.

Most investigations for new physics at long or short  baseline neutrino experiments 
have focussed on measurements made using the charged current (CC) channels, 
with either $\nu_{\mu}\rightarrow\nu_{e}$ or $\nu_{\mu}\rightarrow\nu_{\mu}$ as the 
underlying probabilities, and a final state electron or muon respectively. Our purpose 
in this paper is to study the potential of neutrino NC events at  such 
experiments to provide a tool to investigate features of new physics scenarios. 
This category of neutrino interactions in a typical detector fed by a neutrino beam 
generated in an accelerator facility can comprise neutrino-nucleon and neutrino-electron elastic scattering, neutrino deep-inelastic scattering, 
neutrino-nucleon resonant scattering with a pion in the final state, and finally, neutrino coherent pion scattering\footnote{NC resonant pion scattering in DUNE can comprise the processes $\nu_{\mu}p\rightarrow\nu_{\mu}p\pi^0(n\pi^+)$ and  $\nu_{\mu}n\rightarrow\nu_{\mu}n\pi^0(p\pi^-)$. NC coherent pion scattering from a target nucleus A is the process $\nu_\mu A\rightarrow\nu_\mu A \pi^0$.  }. Similar processes exist, of course, for anti-neutrinos. The relative contributions 
from these various channels  depend on the detector medium, the cross section 
and the energy of the beam, among other things.

\begin{figure}[h]
\center
\includegraphics[width=0.49\textwidth]{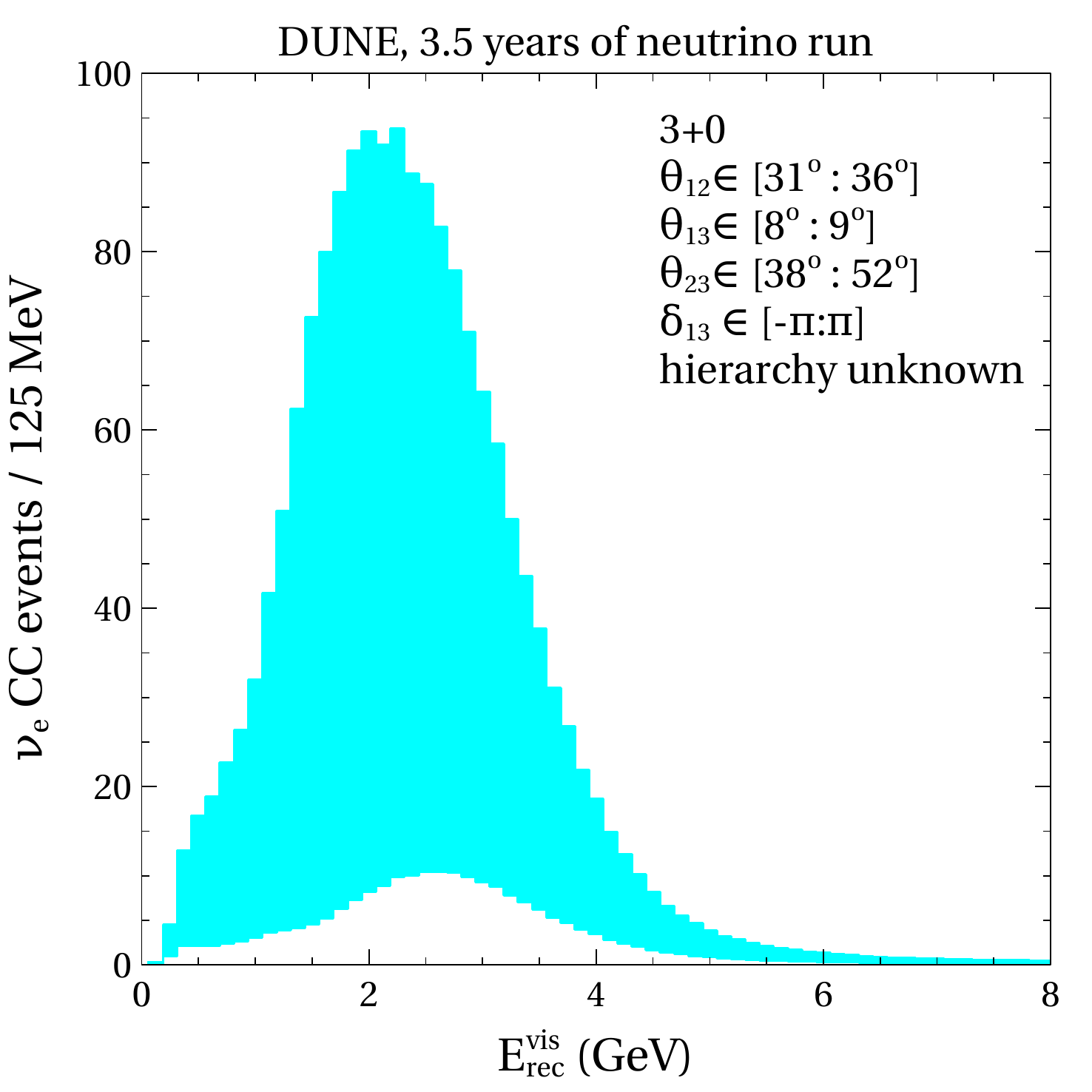}
\includegraphics[width=0.49\textwidth]{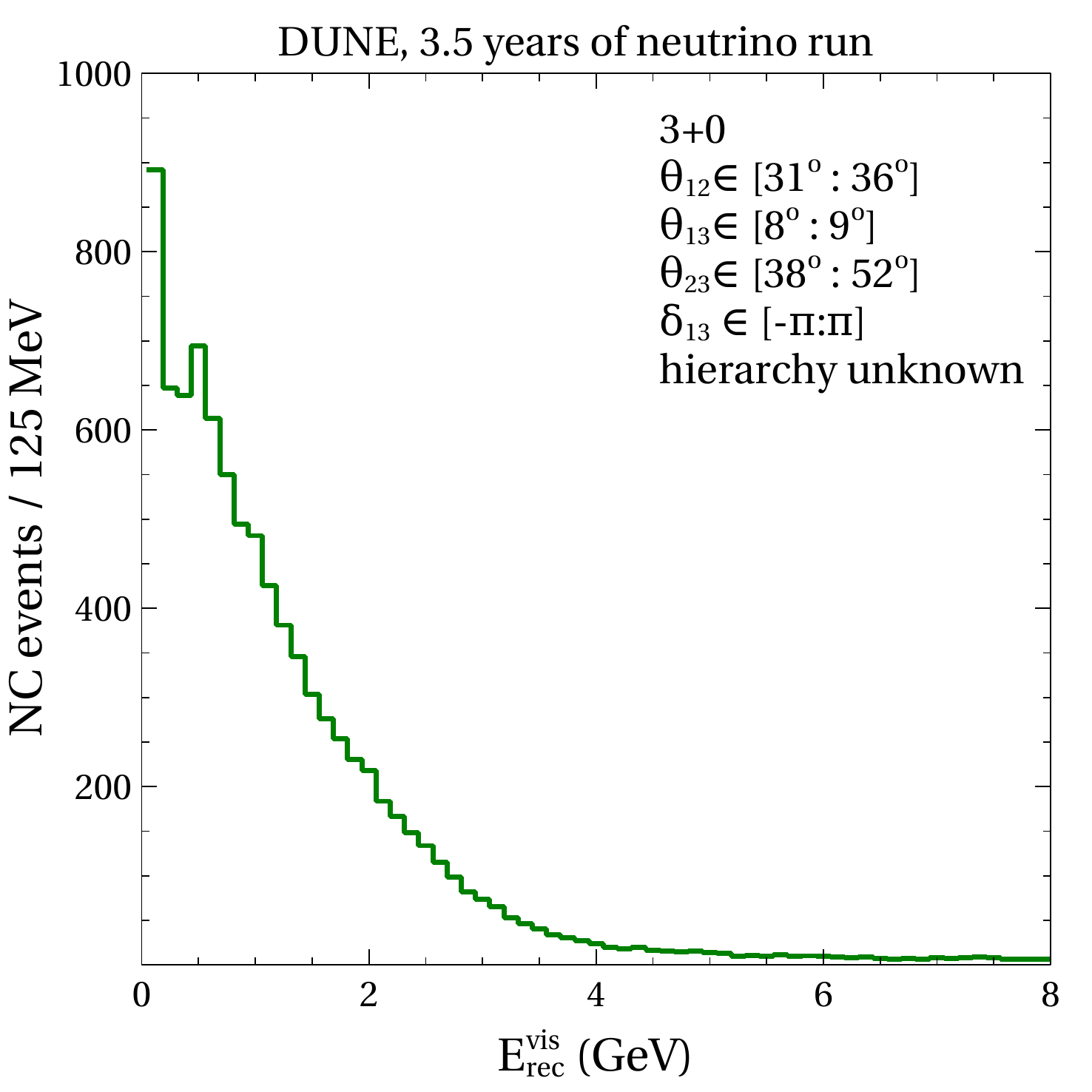}
\caption{\footnotesize{The left and right panel correspond to $\nu_e$ CC and NC 
events respectively. Here, all parameters are varied in their currently allowed range. 
$\rm{E^{vis}_{rec}}$ represents the reconstructed visible energy of the events 
in the detector. In the case of CC events, it closely matches the true energy of the incoming  
neutrino. For the NC events, $\rm{E^{vis}_{rec}}$ can be very different from the 
true incoming neutrino energy, as we discuss later.}}
\label{3+0-uncertainty}
\end{figure}

As we shall show in the remainder of the paper, the measurement and study of 
NC events can in some cases provide a  qualitatively different, complementary and statistically superior
 handle on neutrino properties  in new physics scenarios 
compared to CC measurements. Even when making measurements without 
the presumption of any new physics,  these differences and complementarity 
can be useful. To see this, we consider Fig. \ref{3+0-uncertainty}, which assumes 
the 3+0 scenario. This figure shows, in the left panel, 
the full $3\sigma$ allowed possible band of CC electron events 
at the DUNE far detector given our present knowledge of three generation 
neutrino parameters\footnote{
Throughout this work, we have used the GLoBES software package 
\cite{Huber:2004ka, Huber:2007ji}
along with the snu.c routine 
\cite{jkopp_snu, Kopp:2007ne}
to generate probability, events, and to do $\dxx$-level analyses.}.
The hierarchy  is treated as being unknown, and the 
mixing angles and the CP phase  are varied in their presently allowed ranges. 
Clearly, any measurement by DUNE in this large band is currently acceptable as being consistent with the standard model with massive neutrinos,  
given the present three flavour parameter ranges. The right panel shows the 
NC events for the same parameter variations\footnote{As 
explained in Sec. \ref{NCatLBL}, we use migration matrices provided to us by Michel Sorel to relate reconstructed visible energy in NC events in DUNE to true energy. We note that in the reconstructed energy spectrum in Fig. \ref{3+0-uncertainty}, and also later in the paper in the right panel of Fig. \ref{fig-LE_CC_DUNEevents}, there is a small but somewhat surprising dip between 200 and 300 MeV. We thank Dr. Sorel for checking that this dip is indeed produced by the migration matrices. Since these matrices incorporate many physics details, it is difficult to pin down the origin of the dip more precisely. However, our conclusions are not affected by this dip.}. Besides the superior statistics, 
we note the lack of any dependence on the parameter uncertainties. The reason for this is, of course, the fact that the NC rate is insensitive to any flavour oscillations given the universality of weak interactions. A significant 
deviation from the rate shown, if detected, would clearly indicate the presence of certain kinds 
of new physics (as we discuss later in the paper), as opposed to the CC rate which 
is encumbered by significant uncertainty as well as the possibility of degeneracy between new and 
standard physics.

In subsequent sections, we first emphasise and bring out  a general property of CC versus NC event 
measurements  which can be useful in new physics settings with CP violation 
at both long and short baselines. We show, in a general way,  that NC and CC measurements  complement each other in providing information on CP phases and mixing angles. Then, using the 3+1 scenario\footnote{This is 
dictated less by a belief in the veracity of 3+1 as nature's choice of physics 
beyond the standard model and more by the fact that it offers a simple template 
enabling us to bring out features and draw conclusions which may have applicability 
to other more complex new physics scenarios. Indeed, recent constraints restrict the 
allowed 3+1 parameter space significantly, as we discuss in Sec. \ref{NCatLBL}.} at 
DUNE as an exemplar, we derive an approximate analytic expression for the 
probability governing NC event rates in vacuum , and discuss its features. We find that the 
effects of matter on NC event rates are small, allowing us to use the vacuum expression to good effect. 

We find that NC measurements can be revealing in several ways; for instance, we show that some CP-violating phase 
combinations lead to significant effects on the neutrino and anti-neutrino NC 
probabilities, although not to a significant CP-violating difference between them. 
Nevertheless, we find that under some circumstances there is good sensitivity to these phases. 
We provide bi-probability plots of neutrino versus anti-neutrino NC probabilities for fixed 
mixing angles to show that the CP phases can have substantial effects. We discuss 
how NC events break the degeneracy present in CC events, allowing us to discriminate 
new physics associated with new sterile states from that associated with non-standard 
interactions in neutrino propagation. We identify the general category of new physics scenarios which lend themselves to 
such degeneracy breaking via NC events. Using the 3+1 scenario as an example,  we show the efficacy of NC events in 
constraining  parameters and discuss how they can help improve 
existing bounds. 

Finally, it bears noting that since all three flavours contribute, NC event measurements are typically statistically rich. 
For instance, in the Deep Underground Neutrino Experiment (DUNE), a 7-ton fine-grained 
tracking near detector at $\sim500$ m is planned, and it is expected to detect 
in excess of 400000 NC current events in a year \cite{Acciarri:2015uup}. Similar 
considerations would hold for the planned Short-Baseline Neutrino (SBN) 
program at Fermilab \cite{Camilleri,Antonello:2015lea}. Even at long baselines, NC events are 
typically higher in number compared to any one 
measured CC channel, which buttresses the significance of any conclusions based on their measurement.

\section{Neutral current events in new physics scenarios with CP violation: a general property}

This section identifies a salient property of the NC probability, $P_{NC}$, 
defined as $\Sigma_\beta \pab$, $\alpha,\beta=e,\mu,\tau$  for a given neutrino source beam of flavour $\alpha$ 
and an assumed physics scenario, which will actively contribute to the measured NC rate. 
In the standard $3+0$ scenario, for instance, given a source beam of 
primarily muon neutrinos and  the universality of weak interactions,  
$P_{NC}= P_{\mu e} + P_{\mu\tau}+P_{\mu\mu}=1$. For the same source beam, 
but an assumed 3+1 scenario, 
$P_{NC}=1- P_{\mu s}=P_{\mu e} + P_{\mu\tau}+P_{\mu\mu} \neq 1$, 
where $s$ denotes the sterile flavour. This section is focussed on bringing 
out a feature of NC events that is generic to new physics scenarios with 
CP violation, assuming a 3+2 scenario at a short baseline as an example.

In general, useful conclusions regarding the properties of  $P_{NC}$ can  be drawn by 
examining analytic expressions and comparing them to expressions for  their 
corresponding CC counterparts, $\eg$ $P_{\mu e}$. We begin by 
writing down a general expression for the flavour oscillation probability in vacuum,
\beqa \label{eq:pab_gen}
\nonumber \pab &=& \delta_{\alpha\beta}-4Re\sum_{k>j}(U^{*}_{\alpha k}U_{\beta k}
U_{\alpha j}U^{*}_{\beta j})\sin^2\Delta_{kj}\\
&+& 2Im\sum_{k>j}(U^{*}_{\alpha k}U_{\beta k}U_{\alpha j}U^{*}_{\beta j})\sin2\Delta_{kj}.
\eeqa
Here $k,j$ run over the mass eigenstates, wheras $\alpha,\beta$ denote flavours. Additionally, 
$\Delta_{kj} = 1.27 \times \delta m^{2}_{kj}[\textrm{eV}^{2}] \times L[km]/E[GeV]$ where  $L$ is the baseline
length and $E$ is the neutrino energy.
Eq. \ref{eq:pab_gen} is valid for any number of flavours (including sterile ones, if 
present). 

Consider an experiment sourced by an accelerator generated $\nu_{\mu}$ beam,  
and a 3+2 scenario, with two additional sterile flavour states $\nu_{s_1}$ and $
\nu_{s_2}$, and mass eigenstates $\nu_4$ and $\nu_5$. From Eq \ref{eq:pab_gen}, we 
see that the CP violating part of $P_{\mu s_1}$ resides in 
\beqa \label{eq:pmus_1cp}
\nonumber P_{\mu s_1}^{CP} &\propto& Im\sum_{k>j}(U^{*}_{\mu k}U_{s_1 k}
U_{\mu j}U^{*}_{s_1 j})\sin2\Delta_{kj}\\
\nonumber &\simeq& Im\big[U^{*}_{\mu 5}U_{s_1 5}\big(U_{\mu 4}U^{*}_{s_1 4}
\sin2\Delta_{ 54} + U_{\mu 3}U^{*}_{s_1 3}\sin2\Delta_{ 53} + U_{\mu 2}U^{*}_{s_1 
2}\sin2\Delta_{ 52} + U_{\mu 1}U^{*}_{s_1 1}\sin2\Delta_{ 51}\big) \\
\nonumber &+& U^{*}_{\mu 4}U_{s_1 4}\big(U_{\mu 3}U^{*}_{s_1 3}\sin2\Delta_{ 43} 
+ U_{\mu 2}U^{*}_{s_1 2}\sin2\Delta_{ 42} + U_{\mu 1}U^{*}_{s_1 
1}\sin2\Delta_{ 41}\big)\\
\nonumber &+& U^{*}_{\mu 3}U_{s_1 3}\big(U_{\mu 2}U^{*}_{s_1 2}\sin2\Delta_{ 32} 
+ U_{\mu 1}U^{*}_{s_1 1}\sin2\Delta_{ 31}\big) + U^{*}_{\mu 2}U_{s_1 
2}\big(U_{\mu 1}U^{*}_{s_1 1}\sin2\Delta_{ 21}\big)\big]\\
\eeqa
 A similar expression can be written down for 
$P_{\mu s_2}^{CP}$ with $s_1$ replaced by $s_2$ everywhere, leading to  $$P_{NC} = 
1- P_{\mu s_1}-P_{\mu s_2}.$$

For a short baseline (SBL) experiment, the terms proportional to $\sin2\Delta_{ij}$, 
with $i,j=1,2,3$ can be dropped in comparison to the others, and $P_{\mu s_{1,2}}$ 
simplify; for instance,
\beqa \label{eq:pmus_1cp}
\nonumber P_{\mu s_1}^{CP} &\simeq & Im\big[U^{*}_{\mu 5}U_{s_1 5}\big(U_{\mu 4}U^{*}_{s_1 4}
\sin2\Delta_{ 54} \\
\nonumber &+& U_{\mu 3}U^{*}_{s_13}\sin2\Delta_{ 53} + U_{\mu 2}U^{*}_{s_1 2}
\sin2\Delta_{ 52} + U_{\mu 1}U^{*}_{s_1 1}\sin2\Delta_{ 51}\big) \\
\nonumber &+& U^{*}_{\mu 4}U_{s_1 4}\big(U_{\mu 3}U^{*}_{s_1 3}\sin2\Delta_{ 43} + 
U_{\mu 2}U^{*}_{s_1 2}\sin2\Delta_{ 42} + U_{\mu 1}U^{*}_{s_1 
1}\sin2\Delta_{ 41}\big)\big].\\
\eeqa

For this scenario, the CP violating part of the CC probability, under the 
same approximation as Eq. \ref{eq:pmus_1cp}, is proportional to
\beqa \label{eq:pmue_cp}
\nonumber P_{\mu e}^{CP} &\propto& Im\sum_{k>j}(U^{*}_{\mu k}U_{ek}U_{\mu j}U^{*}_{ej})
\sin2\Delta_{kj}\\
\nonumber &\simeq& Im\big[U^{*}_{\mu 5}U_{e 5}\big(U_{\mu 4}U^{*}_{e4}\sin2\Delta_{ 54}
 + U_{\mu 3}U^{*}_{e 3}\sin2\Delta_{ 53} + U_{\mu 2}U^{*}_{e 
2}\sin2\Delta_{ 52} + U_{\mu 1}U^{*}_{e 1}\sin2\Delta_{ 51}\big) \\
 &+& U^{*}_{\mu 4}U_{e 4}\big(U_{\mu 3}U^{*}_{e 3}\sin2\Delta_{ 43} + 
U_{\mu 2}U^{*}_{e 2}\sin2\Delta_{ 42} + U_{\mu 1}U^{*}_{e 1}\sin2\Delta_{ 41}\big)\big].
\eeqa

In a scenario geared towards explaining the short baseline 
anomalies \cite{Aguilar:2001ty,AguilarArevalo:2008rc,Mention:2011rk,Mueller:2011nm,Aguilar-Arevalo:2013pmq}, further 
simplifications are possible, $\eg$  $\delta m^2_{lm}>>\delta m^2_{mn}$, $l=4$ or $l=5$, $m,n=1,2,3$ \footnote{We stress that the general conclusion we draw in this section remains unchanged with or 
without such simplifications.}. After a little algebra, one then finds that the CP violating difference between NC events measured using an initially muon-flavoured neutrino beam, and those measured using its anti-neutrino counterpart, will be proportional to the quantity $D_{NC}$, given by
\beqa \label{eq:pmu_cp_nc}
D_{NC} &\propto& Im\big[U^{*}_{\mu 5}U_{\mu4}(U_{s_15}U^{*}_{s_14}+U_{s_25}U^{*}_{s_24})\big]
\sin\Delta_{54}\sin\Delta_{43}\sin\Delta_{53}.
\eeqa
On the other hand, the analogous difference for CC events from $\nu_\mu\rightarrow\nu_e$ transitions is proportional to 
\beqa \label{eq:pmu_cp_cc}
D_{CC} &\propto& Im\big[U^{*}_{\mu 5}U_{\mu4}U_{e5}U^{*}_{e4}\big]
\sin\Delta_{54}\sin\Delta_{43}\sin\Delta_{53}.
\eeqa
Comparing Eq. \ref{eq:pmus_1cp} with Eq. \ref{eq:pmue_cp} and Eq. \ref{eq:pmu_cp_nc} with Eq. \ref{eq:pmu_cp_cc}, we see that in both cases  they tap 
into different CP phases and sectors of the mixing matrix. Consequently, the NC 
measurements will provide a qualitatively and quantitatively different window into 
the CP violating and mixing sectors of a new physics scenario compared to the CC 
measurements.   Should a new physics scenario with CP 
violation be nature's choice, then combining NC measurements with CC 
measurements would provide a valuable way to probe it.

\section{Neutral current and new physics at long baselines}
\label{NCatLBL}

For the remainder of this paper, we focus largely, but not exclusively,  on the 3+1 
scenario in order to study the potential of NC events as a probe and diagnostic tool for new physics. 

Additionally, we perform our calculations for the DUNE far detector. 
DUNE \cite{Acciarri:2015uup} is a proposed 
future super-beam experiment with the main aim of
establishing or refuting the existence of CPV in the 
leptonic sector. In addition to this primary goal, this facility will also be
able to resolve the other important issues like the mass hierarchy and 
the octant of $\tz$. The ${\nu_{\mu}(\bar{\nu}_{\mu}})$ super-beam 
will originate at the Fermilab. The optimised beam simulation 
assumes a 1.07 MW - 80 GeV proton beam which will deliver 
$1.47\times10^{21}$ protons-on-target (POT) per year. A 40 kt Liquid Argon 
(LAr) far-detector will be placed in the Homestake mine in South Dakota, 1300 km
away. The experiment plans to have a total of 7 years of running, divided equally 
between neutrinos and anti-neutrinos,  corresponding  to a total
exposure of $4.12\times10^{23}$ kt-POT-yr. The complete experimental 
description of the DUNE experiment such as the CC signal and background definitions 
as well as assumptions on the detector efficiencies concerning the 
CC events are from \cite{Alion:2016uaj}. The details regarding the anticipated
NC events at DUNE were taken from \cite{Adams:2013qkq}. The NC event
detection efficiency has been assumed to be $90\%$. In order to correctly
reproduce the NC events spectra, we have made use of the {\it migration matrices}. 
In a NC event, the outgoing (anti-)neutrino carries away some fraction of the 
incoming energy. This energy is missed and hence, the reconstructed visible
energy is less than the total incoming energy. As such, the events due to
energetic (anti-)neutrinos are reconstructed inaccurately 
at lower visible energies in a majority of such cases\footnote{The profile
of NC events spectrum, for example, can be seen in the right panel of Fig. \ref{3+0-uncertainty}.}.
Therefore, using a gaussian energy resolution function in such a situation is not appropriate. 
We have used the migration matrices from \cite{DeRomeri:2016qwo}, 
provided to us by \cite{msorel}. 
Note that these migration matrices correspond to a binning of 50 MeV 
and therefore, in this work too, we have considered the energy bins of 50 MeV
for NC events\footnote{For the sake of clarity, we show NC events in Figs. \ref{3+0-uncertainty} 
and \ref{fig-LE_CC_DUNEevents} in energy bins of 125 MeV. However, 
for binned-$\dxx$ calculations, we have considered 50 MeV energy bins.}. 
For the analysis of CC events, we have used energy bins of
125 MeV as in \cite{Alion:2016uaj}. The background to NC events consists
of CC events that get mis-identified as NC events. These include electron
events (due to CC signal $\nu_{\mu}\rightarrow\nu_{e}$ or intrinsic beam $\nu_{e}\rightarrow\nu_{e}$), 
muon events ($\nu_{\mu}\rightarrow\nu_{\mu}$), tau events ($\nu_{\mu}\rightarrow\nu_{\tau}$)
and their respective CP-reversed channels due to anti-neutrino/neutrino 
contaminations in the beam. It should be noted that the backgrounds too, will 
oscillate into the sterile flavour depending on the values of $U_{e4}, U_{\mu 4}$
and $U_{\tau 4}$. In such a scenario, the simplifying assumptions of 
putting one or more of these matrix elements to 0, {\it may not} give the 
correct estimate of the NC signal events. For the NC analysis, the signal and background 
normalisation errors have been taken to be $5\%$ and $10\%$ respectively.

Finally, we note that in the 3+1 scenario, flavor oscillations may lead to
some depletion of the active neutrino flux and of its muon neutrino component at the location of the DUNE
near detector ($\sim 500$ m). This could, in principle, distort the flux
measurement made at this location, which
forms the basis of conclusions drawn regarding oscillations measured at the
far detector. We have assumed an overall error of $5\%$ in flux
measurements, and have checked that given the currently allowed parameter
ranges for the 3+1 scenario, the change in flux due to a sterile species is
always below this limit. On the other hand, as we show in this work,
depletion in the NC rate significantly above this uncertainty is expected at
the far detector, hence enabling DUNE to detect the possible presence of a
sterile state via neutral current measurements.

\subsection{An approximate analytical expression for $P_{\mu s}$}

As mentioned above, the NC rate in a $3+1$ scenario will be proportional to 
$1-P_{\mu s}$. We give below a useful approximate expression, starting from
Eq. \ref{eq:pab_gen}, since the full expression which follows from it is extremely 
long and complicated.
In obtaining this approximate form, we have adopted the following parameterisation 
for the PMNS matrix: 
\beq
U_{\text{PMNS}}^{3+1}=O(\theta_{34},
\delta_{34})O(\theta_{24},\delta_{24})O(\theta_{14})O(\theta_{23})O(\theta_{13},
\delta_{13})O(\theta_{12})
\eeq
Here,  in general, $O(\theta_{ij},\delta_{ij})$ is a rotation matrix
in the $ij$ sector with associated phase $\delta_{ij}$. For example,

\begin{equation*}
O(\theta_{24},\delta_{24}) = 
\begin{pmatrix}
1 & 0 & 0 & 0 \\
0 & \cos\tb & 0 & e^{-i\delta_{24}}\sin\tb \\
0 & 0 & 1 & 0 \\
0 & -e^{i\delta_{24}}\sin\tb & 0 & \cos\tb
\end{pmatrix};~
O(\theta_{14}) = 
\begin{pmatrix}
\cos\ta & 0 & 0 & \sin\ta \\
0 & 1 & 0 & 0 \\
0 & 0 & 1 & 0 \\
-\sin\ta & 0 & 0 & \cos\ta
\end{pmatrix} \text{\etc}
\end{equation*}

\begin{figure}[h]
\center
\includegraphics[width=0.49\textwidth]{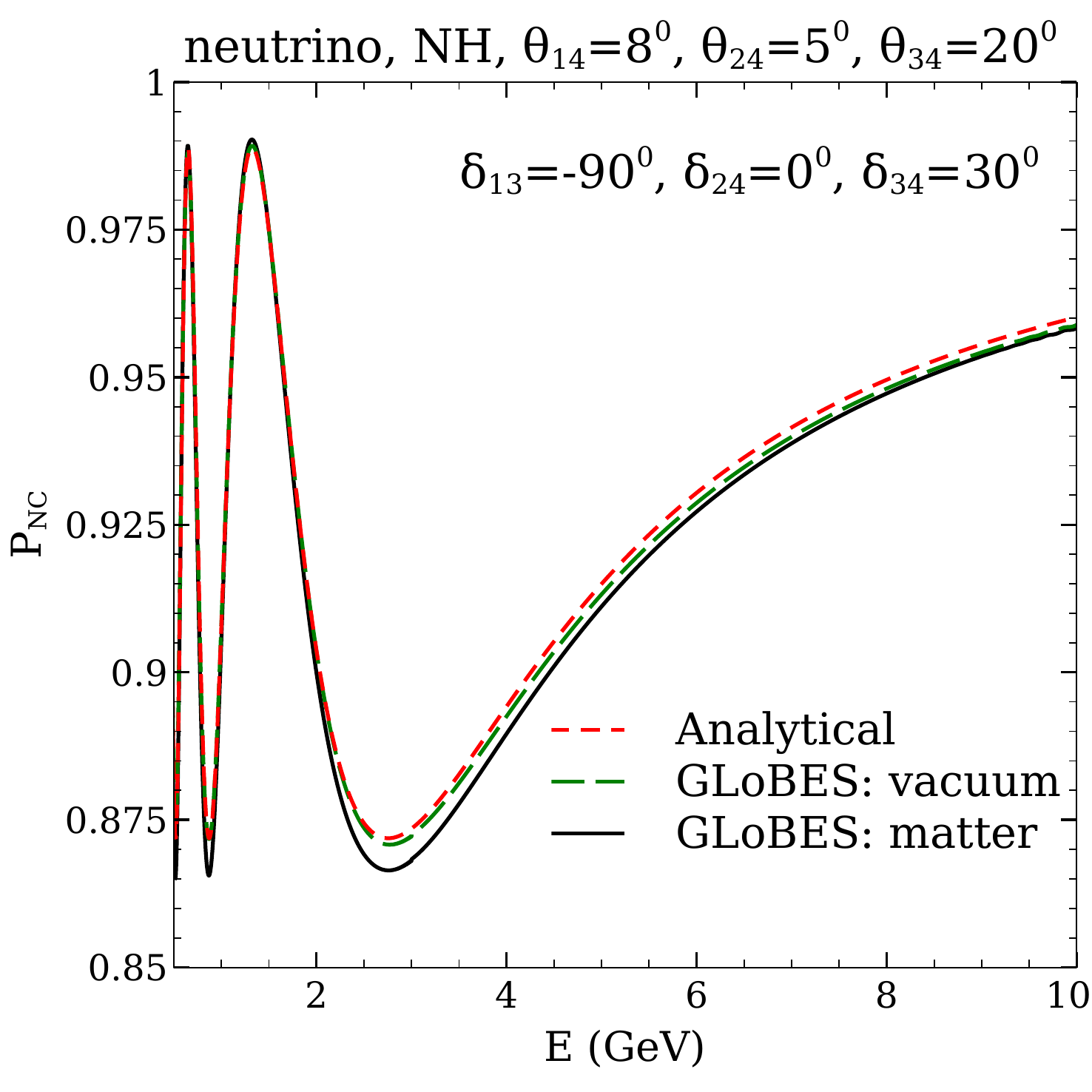}
\includegraphics[width=0.49\textwidth]{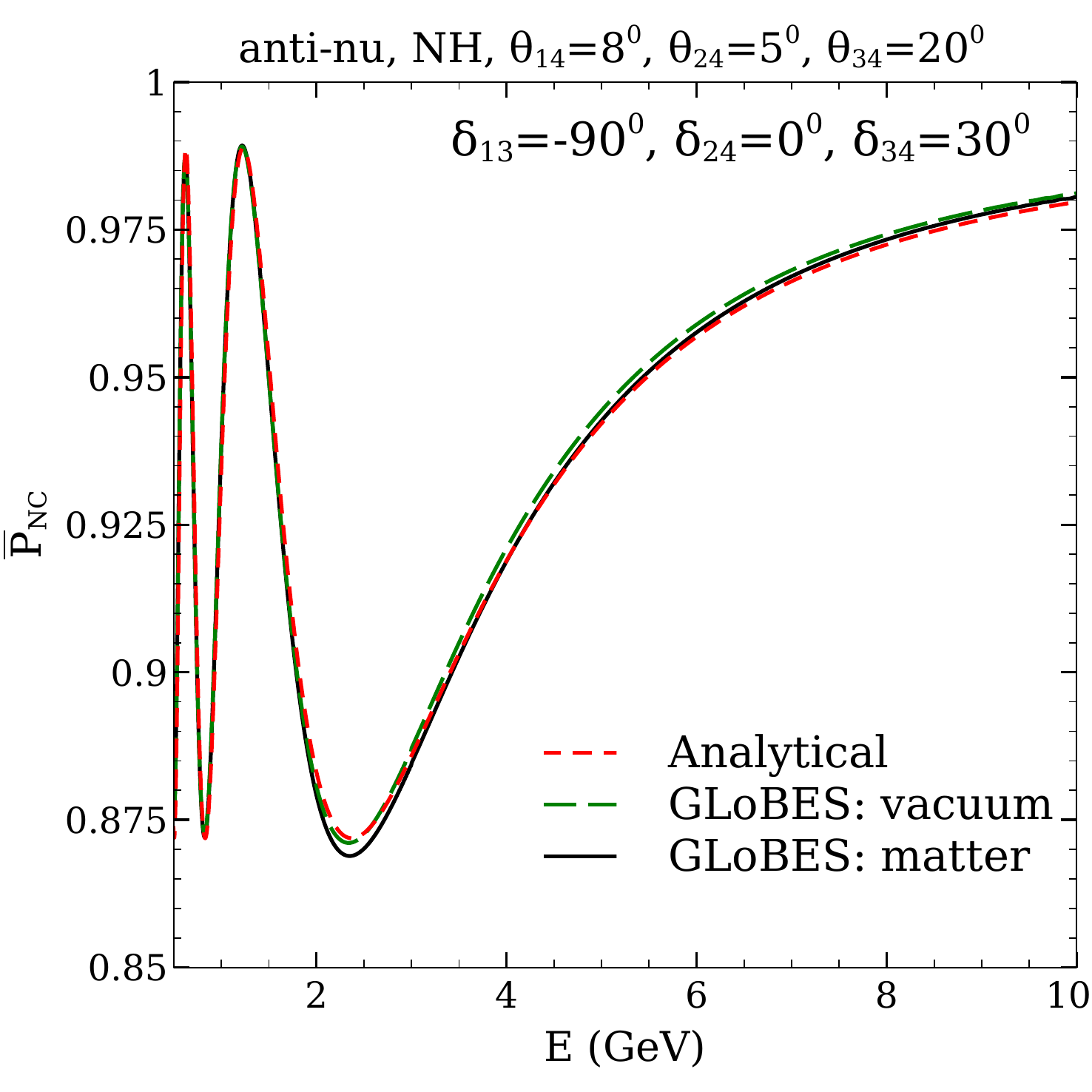}
\caption{\footnotesize{Probability plots , comparing $P_{NC}=1- P_{\mu s}$ using 
the approximate formula for $P_{\mu s}$given in the text, Eq. \ref{eq-pmsvacuum} (red dashed 
curves) with the full GLoBES result in vacuum (green dashed curves) and matter 
(black solid curves). The left panel is for neutrinos and the right one for anti-neutrinos.  
Choices of phases and mixing angles have been made as shown. The curves 
correspond to $L=1300~\rm{km}$.}}
\label{theo_vs_sim_check1}
\end{figure}

Measurements from MINOS, MINOS+, Daya Bay and the IceCube
experiments provide significant constraints on the 3+1 paradigm. See, for instance, 
\cite{Carroll:2017xps, Adamson:2016jku, TheIceCube:2016oqi}. The Super-Kamiokande data, 
MINOS NC data, NO$\nu$A NC data and the IceCube-DeepCore data  
provide constraints on the 3-4 mixing \cite{Abe:2014gda, MINOS:2016viw, Aartsen:2017bap, Adamson:2017zcg}. 
Our work in this paper utilises only the currently allowed parameter space for 
this scenario as determined by these references\footnote{It should be noted that
there are global analyses of the existing oscillation data that provide constraints
on the 3+1 paradigm \cite{Kopp:2013vaa, Collin:2016aqd, Gariazzo:2017fdh}. 
However, there exist differences in their  results corresponding
to the fits in the parameter space $\Delta m^2_{41}-\sin^2\tc$. There is also 
the difficulty in reconciling the appearance data with the disappearance data. 
Keeping these points in mind, we adhere to the constraints on 3+1 from the disappearance
data from the above-mentioned standalone experiments.}. 
Since these current constraints restrict  $\theta_{13} , \theta_{14} , 
\theta_{24} \leq 13^\circ$, we take $\sin^3\theta_{ij}=0$, where $
\theta_{ij}$ is any of these angles. We also set  $\theta_{23}=45^\circ$ for simplicity, 
and assume $\sin^2\dfrac{\Delta m^2_{31}L}{4E}=\sin^2\dfrac{\Delta m^2_{32}L}{4E}$, 
while neglecting the contribution from the solar mass-squared difference,  
since $\Delta m^{2}_{21} << \Delta m^{2}_{31}$. Additionally, we work under the 
assumption that the mass-squared differences $\delta m^2_{lm}$, $l=4$, $m=1,2,3$ are all 
approximately equal, implying that the fourth mass eigenstate is much heavier than 
the other three. With these simplifications, we obtain, for the vacuum transition 
probability for $\nu_{\mu}$ to $\nu_{s}$,
\begin{eqnarray}
\label{eq-pmsvacuum}
\rm P^{vac}_{\rm{\mu}s}  &\simeq&  \cos^{4}\theta_{14} \cos^{2} \theta_{34} 
\sin^{2} 2\theta_{24}\sin^2\dfrac{\Delta m^2_{41}L}{4E} \nonumber \\ &+&  \Big[\cos^4\theta_{13} 
\cos^2\theta_{24} \sin^2\theta_{34}  - \cos^2\theta_{13} \cos^2\theta_{24} 
\cos^2\theta_{34} \sin^2\theta_{24} \nonumber \\ 
&+& \dfrac{1}{\sqrt{2}} \sin2\theta_{13} \sin2\theta_{34}\sin\theta_{14} \cos^{3}\theta_{24} 
\cos(\delta_{13} + \delta_{34})\Big] \sin^2\dfrac{\Delta m^2_{31}L}{4E} \nonumber \\
&+&\dfrac{1}{2}\cos^2\theta_{13} \cos^2\theta_{24}\sin2\theta_{34} \sin\theta_{24} 
\sin (\delta_{34}-\delta_{24})\sin\dfrac{\Delta m^2_{31}L}{2E}.
\end{eqnarray}
Prior to testing the accuracy of this formula and determining its applicability, we 
note the following characteristics:
\begin{enumerate}
\item The first term is, in its exact form, a rapidly oscillating term due to the large 
mass-squared difference.  In the plots to follow, for specificity, we assume it to be  $
\simeq 1$ eV$^2$, and  adopt the DUNE baseline of $1300$ km.

\item Of the three phases, only two linear combinations  appear: 
$\delta_1=  \delta_{13} + \delta_{34}~\rm{and}~\delta_2=  \delta_{34} - \delta_{24}$; 
and only the latter 
is responsible for CP violation in neutral currents.
\end{enumerate}
It follows that these simplifications and characteristics percolate into 
$P_{NC}=1- P_{\mu s}$, which we now plot in Fig. \ref{theo_vs_sim_check1}. 

We see that there is good agreement between the exact GLoBES curves 
(solid black  and dashed green lines) and the ones generated by the analytical 
approximation (red dashed line). The curves  show no rapid oscillations since the small wavelength oscillation part of
first term in $P_{\mu s}$ is averaged out to $0.5$.

\begin{figure}[h]
\center
\includegraphics[width=0.49\textwidth]{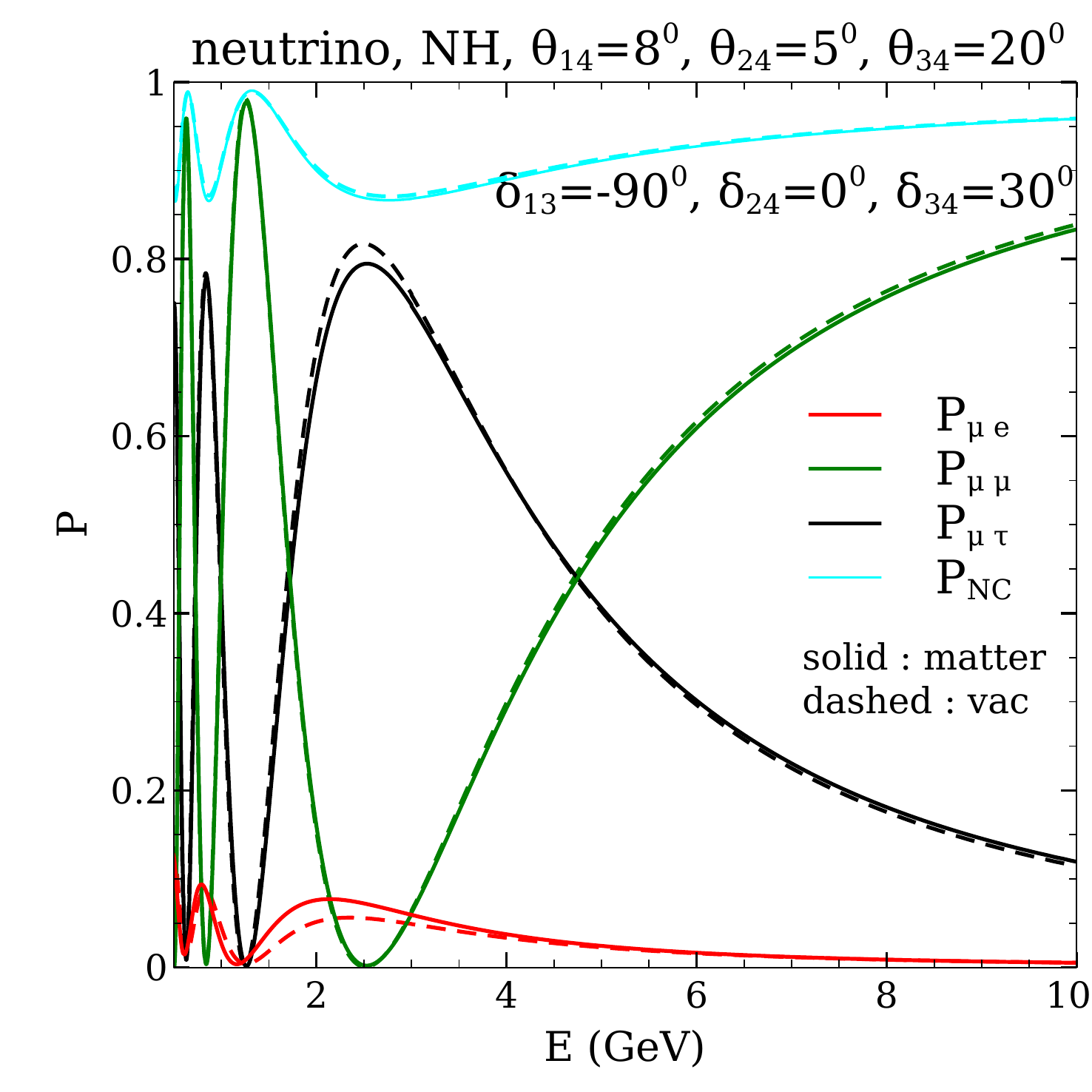}
\includegraphics[width=0.49\textwidth]{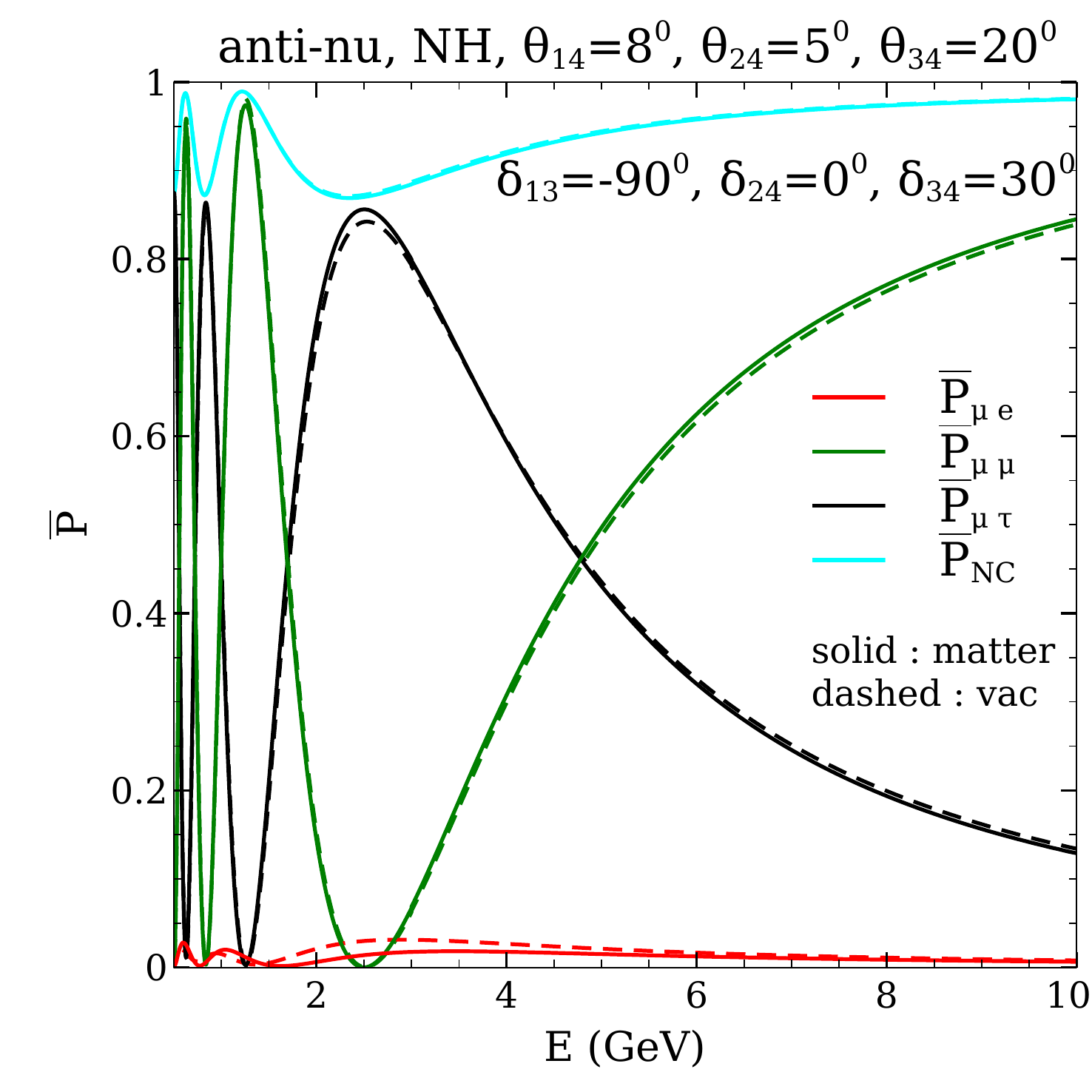}
\caption{\footnotesize{Probability plots , comparing $P_{\mu e}, P_{\mu\tau}, 
P_{\mu\mu}$ and $P_{NC}=1- P_{\mu s}$, all for the 3+1 model,  using GLoBES. 
The left 
panel is for neutrinos and the right one for anti-neutrinos, and solid curves are for 
matter while the dashed ones are for vacuum.  Choices of phases and mixing angles 
have been made as shown. The curves 
correspond to $L=1300~\rm{km}$.}}
\label{all_nu_antinu}
\end{figure}
 From Fig. \ref{theo_vs_sim_check1} we see that the approximate formula, derived 
 for the vacuum case,  also works well for matter, $\ie$  the overall matter effect in NC 
event rates is small. Some understanding of this feature  can be gleaned from 
Fig \ref{all_nu_antinu}, which shows full GLoBES curves for the various probabilities, and 
demonstrates how the $\nu_{\mu}\rightarrow\nu_e$ and $\nu_{\mu}\rightarrow\nu_\tau$ channels 
have matter effects that are already small in each of these channels, and that 
nearly cancel each other over the DUNE energy 
range and baseline. While we certainly cannot generalise this over baselines,  
energies and new physics scenarios, we note that such a near cancellation can  
occur for a range of baselines and energies in the 3+0 scenario\footnote{For a fuller 
discussion of the 3+0 case see \cite{Gandhi:2004bj}.}.

Finally, we note that the NC probabilities for neutrinos and anti-neutrinos are very similar, 
as a comparison of the left and right panels in Figs. \ref{theo_vs_sim_check1} 
and \ref{all_nu_antinu} demonstrate.

\subsection{Effect of the CP violating phases on $P_{NC}$ }

This section attempts to understand the dependence of $P_{NC}$ on the
three CP violating phases $\da,\db$ and $\dc$ in a simple way.

\begin{figure}[h]
\center
\includegraphics[width=0.49\textwidth]{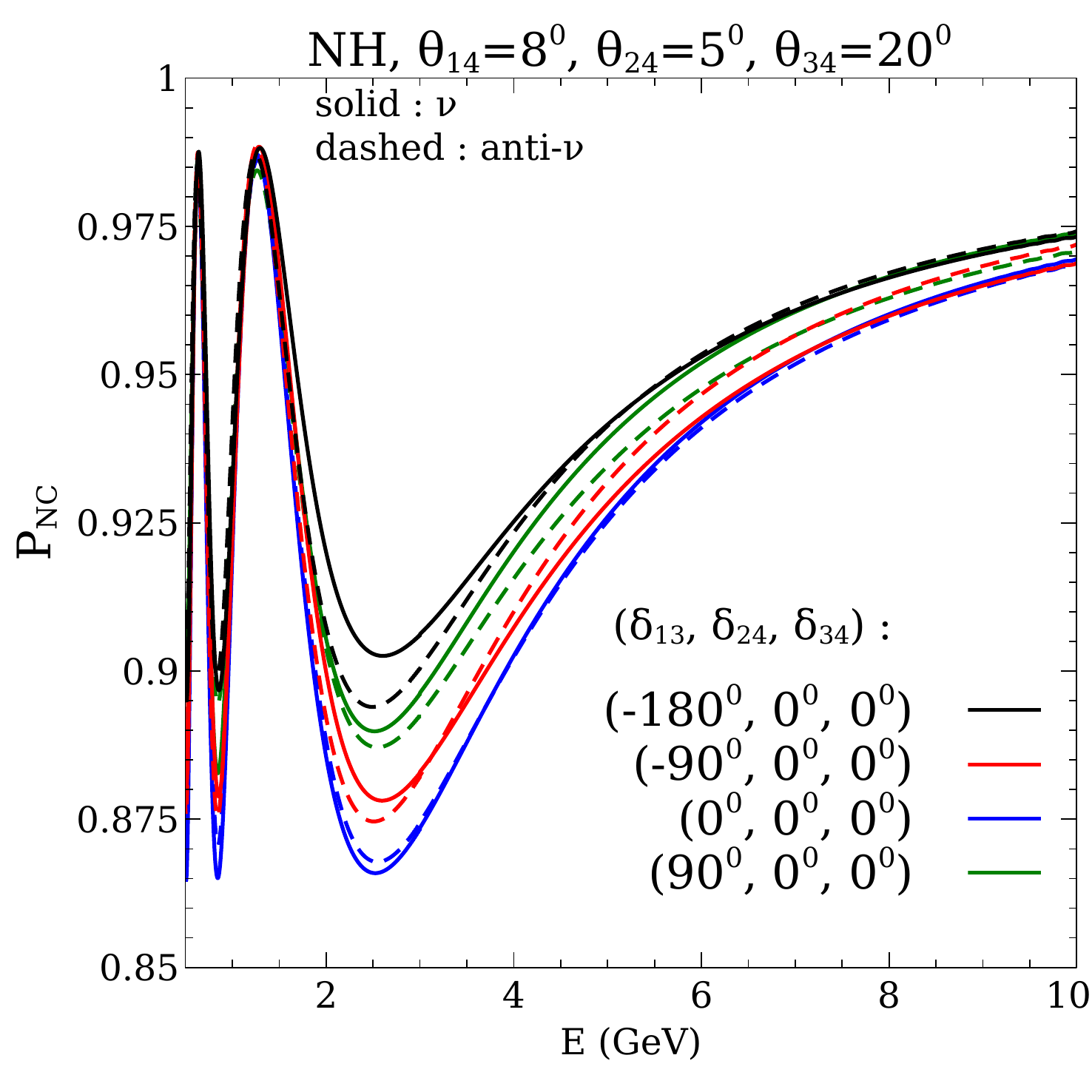}
\includegraphics[width=0.49\textwidth]{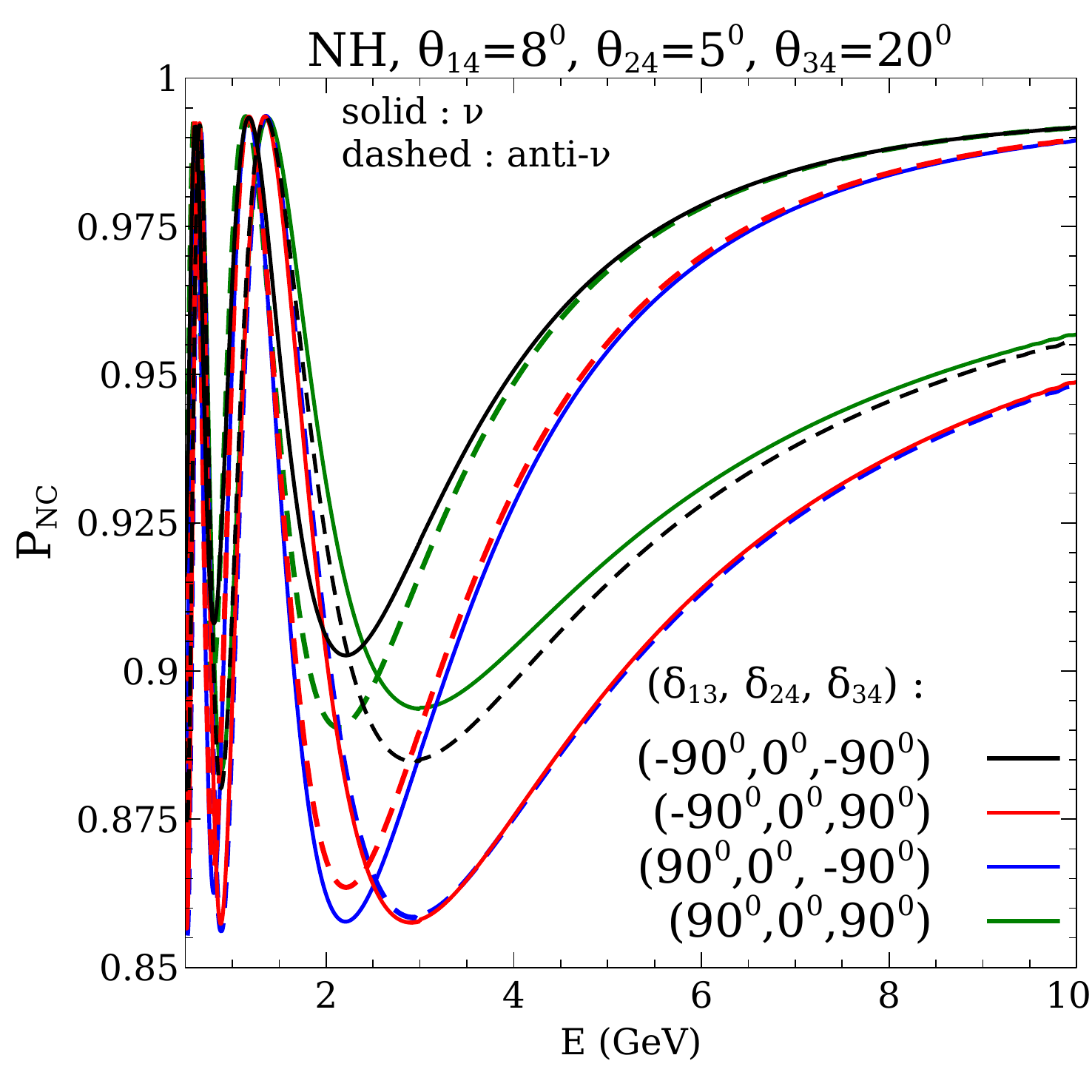}
\caption{\footnotesize{$P_{NC}$ vs Energy (in GeV) assuming normal hierarchy 
for $L=1300~\rm{km}$ in the presence of matter. 
$\ta = 8^\circ, \tb = 5^\circ, \tc=20^\circ$ (fixed). The solid (dashed) curves
are for neutrino (anti-neutrino). Left panel: $\da \in \{-180^\circ,-90^\circ,0,90^\circ\}$, 
$\db = \dc = 0$. Right panel: $\da,\dc \in \{\pm90^\circ,\pm90^\circ\}$ and $\db=0$ 
as shown in the key.}}
\label{fig-CPdependence-1}
\end{figure}

\begin{figure}[h]
\center
\includegraphics[width=0.49\textwidth]{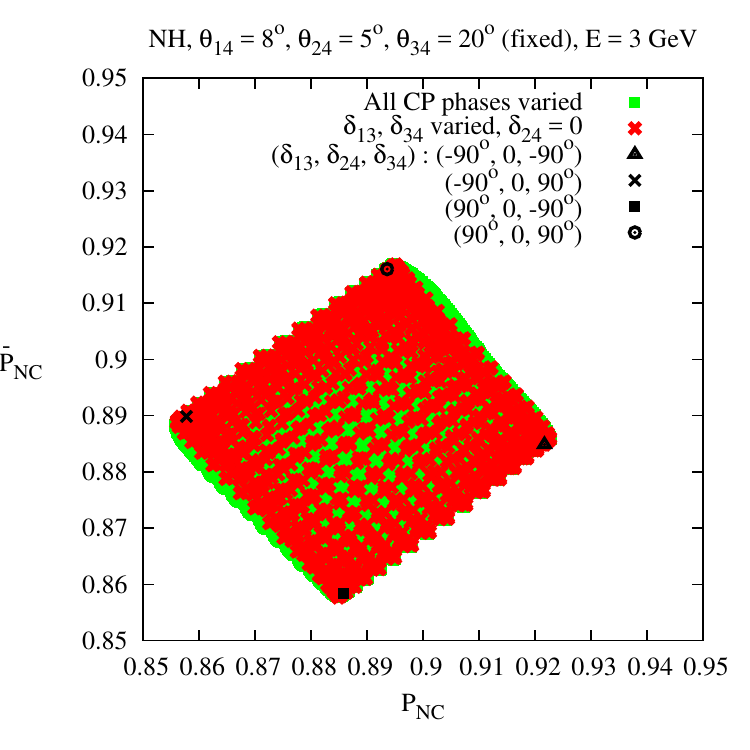}
\includegraphics[width=0.49\textwidth]{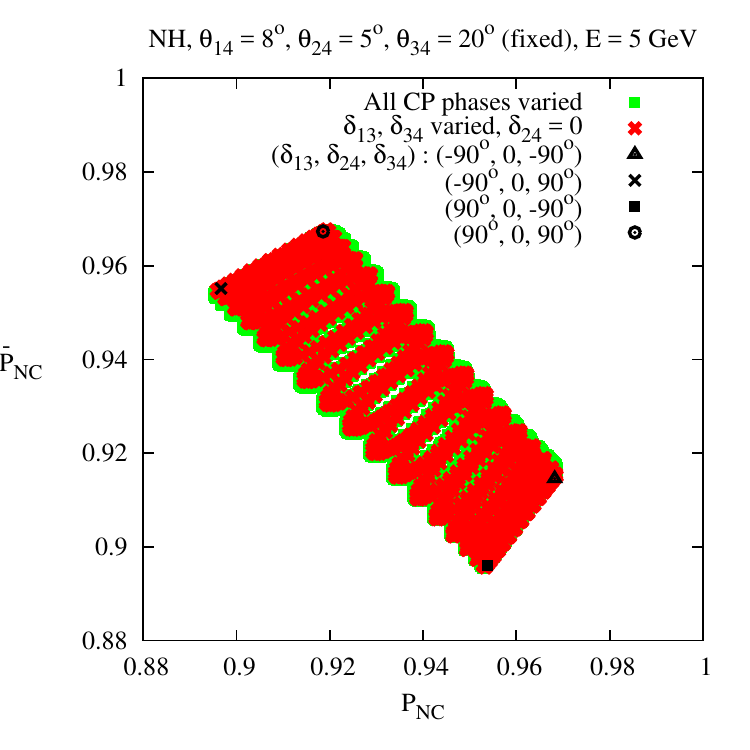}
\caption{\footnotesize{$\bar{P}_{NC}$ vs $P_{NC}$ at $E = 3~GeV$ (left-panel) and at
$E = 5~GeV$ (right-panel) for $L=1300~\rm{km}$ in the presence of matter. 
$\ta = 8^\circ, \tb = 5^\circ, \tc=20^\circ$ (fixed). Green region:
This region corresponds to all $\bar{P}_{NC}$ - $P_{NC}$ values that are obtained 
when all the three CP phases are varied in $[-180^\circ,180^\circ]$. Red region:
This region corresponds to all $\bar{P}_{NC}$ - $P_{NC}$ values that are obtained 
when $\da$ and $\dc$ are varied in $[-180^\circ,180^\circ]$ while $\db=0$. Black points: 
$\da,\dc \in \{\pm90^\circ, \pm90^\circ\}$ when $\db=0$.}}
\label{fig-CPdependence-3}
\end{figure}

In Fig. \ref{fig-CPdependence-1} we have plotted the $P_{NC}=1-P_{\mu s}$ as a function
of the Energy (GeV) in the presence of matter for $L = 1300$ km. The 
plots correspond to normal hierarchy, $\ta=8^\circ$, $\tb=5^\circ$
and $\tc=20^\circ$. The solid curves show the probability values for neutrinos while the
dashed ones are for anti-neutrinos. In the left panel we show the dependence 
of $P_{NC}$ on the $\da$ phase. For this panel,  we show curves corresponding to 
$\da = -180^\circ, -90^\circ, 0 ~\rm{and}~ 90^\circ$. The other two phases 
$\db ~\rm{and}~ \dc$ have been set to 0. In the right panel, we show the 
dependence of $P_{NC}$ on the $\da$ and $\dc$ phases with 
the $\db$ phase  kept equal to 0. We show four set of curves for both
neutrinos and anti-neutrinos corresponding to $\da,\dc \in \{\pm90^\circ, \pm90^\circ\}$.
From Fig. \ref{fig-CPdependence-1}, we can draw the following  conclusions:
\begin{itemize}
\item $P_{NC}$ has significant dependence on the CP phases $\da$ and $\dc$.
\item The left panel shows that the differences 
between neutrino and anti-neutrino probabilities
are small. However, there is appreciable 
separation between the $\da=0,-180^\circ$ 
and $\da=-90^\circ, 90^\circ$curves. This can be 
understood from Eq. \ref{eq-pmsvacuum} 
where the $\da$-dependence is through a cosine term.
\item In the right panel, the  introduction of the  $\dc$ 
phase induces larger differences between the 
neutrino and anti-neutrino probabilities, specially at 
higher energies. Referring to Eq. \ref{eq-pmsvacuum}, we see that as the energy increases, the CP violating term will tend to undergo less suppression compared to the other terms, hence its effect tends to become more visible. Thus, the measurement of a large CP-asymmetry
in the NC events at DUNE can point to a CP-violating value 
of $\dc$\footnote{Or, more accurately,
a CP-violating value of $\delta_2= \dc - \db$, as we emphasise below.}. 
\item In the right panel, for neutrinos, while the peaks for all curves 
have about the same value of $P_{NC}$, among the minima, the lowest 
value of  ${P_{NC }}$ occurs for $(\da,\dc)$ values 
around $(-90^\circ, 90^\circ)$ while the highest value 
 occurs for $(\da,\dc)$ values 
around $(-90^\circ, -90^\circ)$. For anti-neutrinos, again, examining minima,  the lowest value of  $P_{NC }$ occurs for 
$(\da,\dc)$ values around $(90^\circ, -90^\circ)$ while the highest value  occurs 
for $(\da,\dc)$ values around $(90^\circ, 90^\circ)$. This again, is easy to understand
from Eq. \ref{eq-pmsvacuum}, where the CP dependence is of the form 
$A\cos(\da+\dc) + B\sin{\dc}$ for $\db=0$. Note that here we have shown the curves 
for restrictive values of $\da$ and $\dc$, but  this behaviour is verified
again in Fig. \ref{fig-CPdependence-3} in a more general way.
\item It is also evident from the right panel of Fig. \ref{fig-CPdependence-1} 
that the probability curve for neutrino corresponding to $(\da,\dc)$ of let's 
say $(x,y)$ where $x, y = \pm90^\circ$ is degenerate with the anti-neutrino curve of $(-x,-y)$, especially at higher 
and lower energies. Small differences due to matter effects can be seen near the minima. 
Neglecting these small matter effects, we see from the approximate expression for the vacuum oscillation probability, Eq. \ref{eq-pmsvacuum}, that the degenerate probability curves should indeed be identical.

\end{itemize}

\begin{figure}[h]
\center
\includegraphics[width=0.49\textwidth]{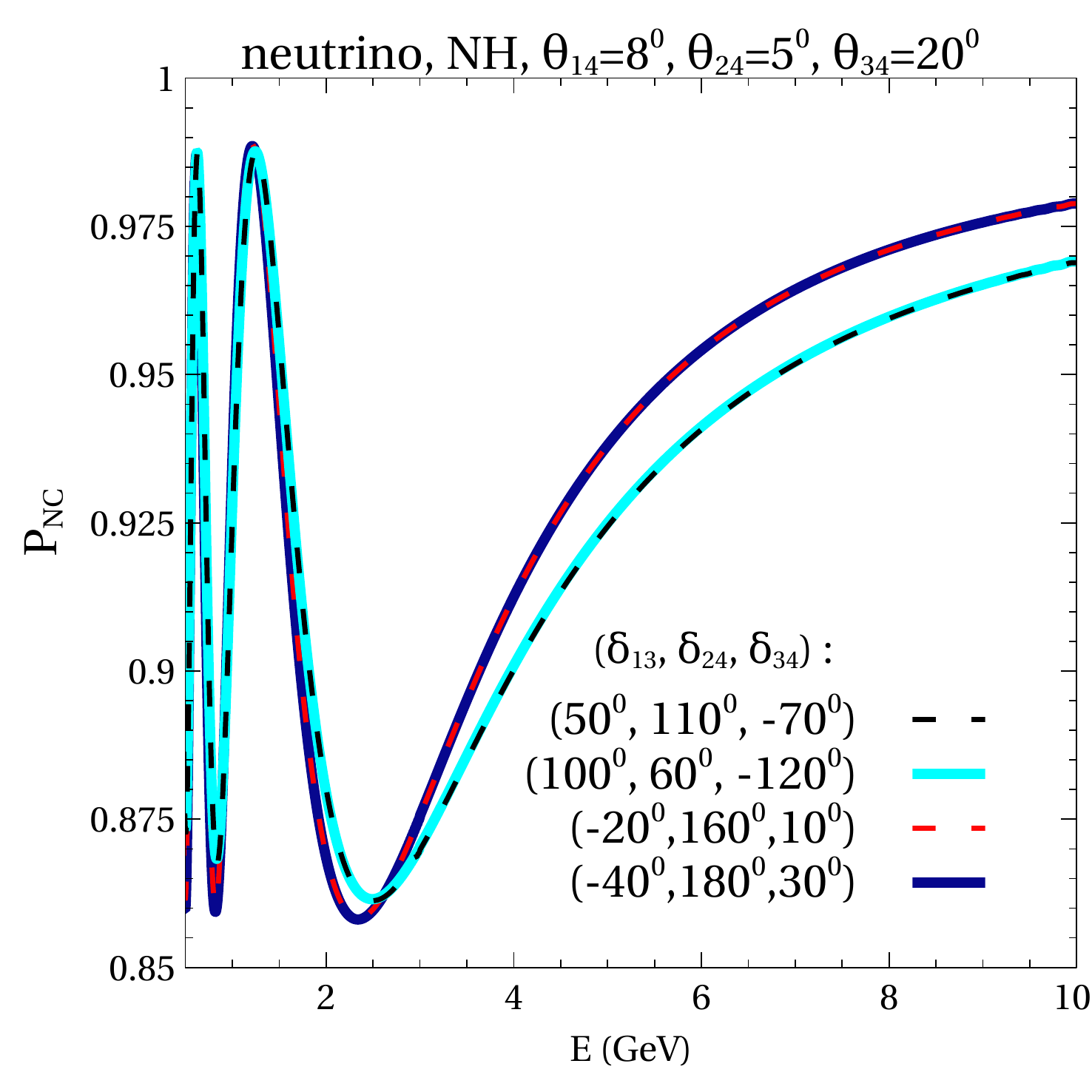}
\includegraphics[width=0.49\textwidth]{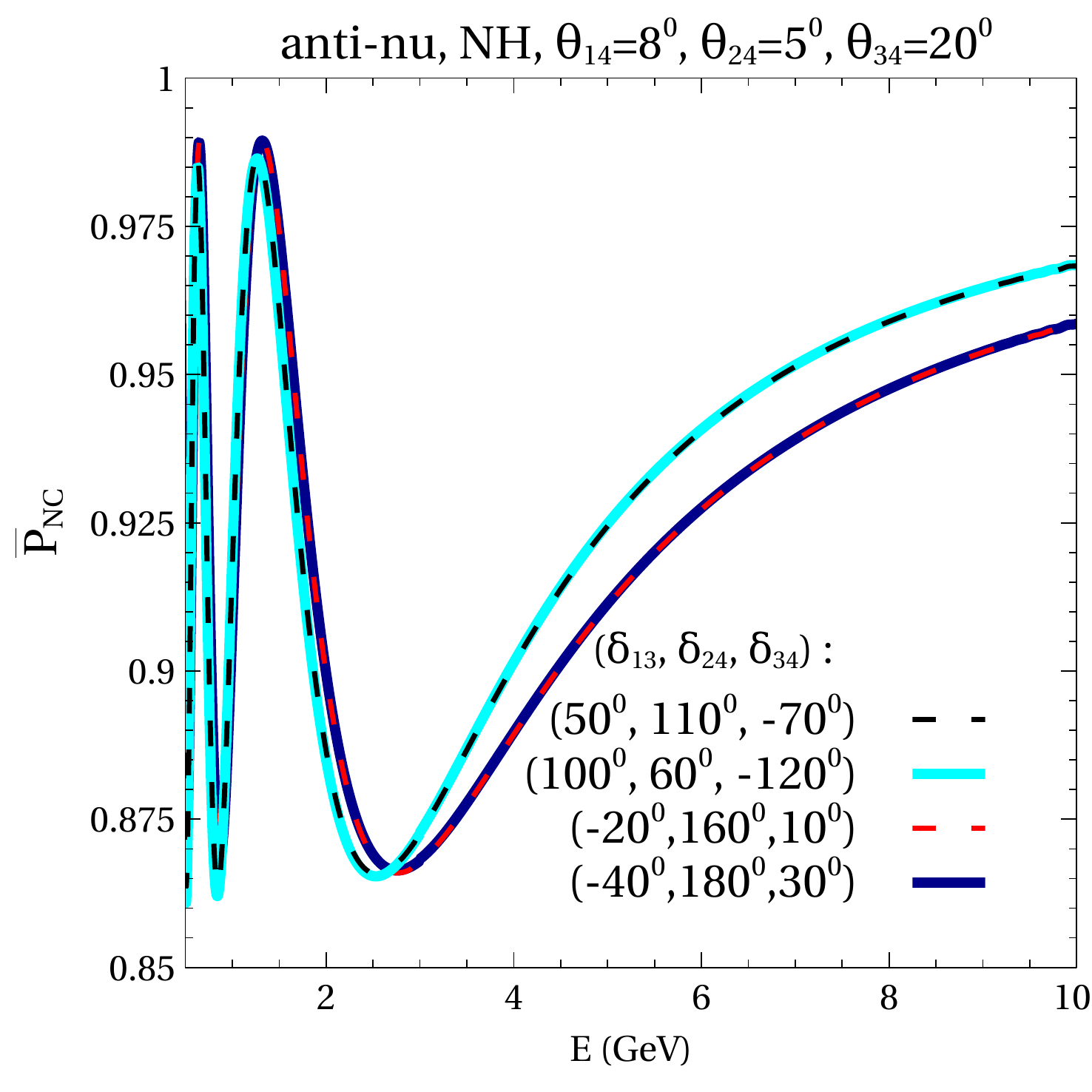}
\caption{\footnotesize{$P_{NC}$ vs Energy (in GeV) assuming normal hierarchy 
for $L=1300~\rm{km}$ in the presence of matter.
$\ta = 8^\circ, \tb = 5^\circ, \tc=20^\circ$ (fixed). The left (right) panel 
corresponds to neutrino (anti-neutrino) probabilities. Different values of 
$\da, \db, \dc$ are chosen as shown in the key.}}
\label{fig-CPdependence-2}
\end{figure}

Under the approximations in which 
Eq. \ref{eq-pmsvacuum} is valid, it can be seen that there are only two effective CP phases
that will play a role in $P_{NC}$ at the leading order. 
These are $\delta_{1} = \da+\dc$ and 
$\delta_{2} = \dc-\db$. Thus, in our chosen parameterisation of the PMNS 
matrix, there is a degeneracy between the three CP phases. We show this
explicitly in Fig. \ref{fig-CPdependence-2}. We have plotted $P_{NC}$ as
a function of the energy for neutrinos (anti-neutrinos) in the left (right) panel.
We choose two sets of different $(\da,\db,\dc)$ values (as shown in the 
key in the figures) which give the same $\delta_{1}$ and $\delta_{2}$ values. 
The assumed values of the  other oscillation parameters are same as Fig. \ref{fig-CPdependence-1}.
It can be seen that there is almost complete degeneracy between the curves corresponding to common values of the phases $\delta_1$ and $\delta_2$.
Note that  Eq. \ref{eq-pmsvacuum} is derived for vacuum, and for the special value $\theta_{23}=45^o$. However the 
degeneracies hold true for matter probabilities at $L=1300$ km and for other assumed values of $\theta_{23}$ within its allowed range.

It is therefore possible to set one of the phases equal to 0, without the loss of generality. 
Since, we have considered $\sin\tb$ to be a small quantity as its range of values 
is the most restricted, putting $\db=0$ may be the best choice in order to not have 
significant differences between vacuum and matter probabilities. We explore this in 
Fig. \ref{fig-CPdependence-3}, generated using GLoBES. These plots show the values of probabilities in 
the $P_{NC}$ - $\bar{P}_{NC}$ plane for different values of the oscillation parameters.
In Fig. \ref{fig-CPdependence-3}, we show results for normal hierarchy, $\ta=8^\circ$, 
$\tb=5^\circ$ and $\tc=20^\circ$. The left (right) panel corresponds to neutrino energy 
of 3 GeV (5 GeV). The green region shows the space in the $P_{NC} - \bar{P}_{NC}$ plane when all the 
three phases are varied in the range $[-180^\circ, 180^\circ]$. The red region corresponds
to the space when $\da$ and $\dc$ are varied in $[-180^\circ, 180^\circ]$, holding 
$\db$ equal to 0. The four black points correspond to $\da,\dc \in \{\pm90^\circ, \pm90^\circ\}$
when $\db=0$. From Fig. \ref{fig-CPdependence-3}, we conclude that 
\begin{itemize}
\item The fact that the red region is almost the same as the green region suggests that 
putting $\db=0$ does not lead to any loss of obtainable $P_{NC}$ - $\bar{P}_{NC}$
space.
\item The four black points - $\da,\dc \in \{\pm90^\circ, \pm90^\circ\}$
indeed quite closely correspond to the $\da, \dc$ values for which the 
the $P_{NC}$ and $\bar{P}_{NC}$ are minimum or maximum.
This is true for both 3 GeV and 5 GeV.
\item The dependence on the $\da$ and $\dc$ phases as expressed  in Eq. \ref{eq-pmsvacuum} is 
reasonably accurate.
\end{itemize}

Finally, we point out that the situation above serves as another example 
of the point made in Section 2. The NC events for the long baseline of 
DUNE and the chosen 3+1 new physics scenario provide a 
window into CP  via the phase combinations $\delta_{1} = \da+\dc$ and 
$\delta_{2} = \dc-\db$ both in vacuum and in matter. On the 
other hand,  as discussed in \cite{Gandhi:2015xza}, the CC probability 
$P_{\mu e}$ in matter  for the same 
scenario is sensitive to all three CP phases\footnote{In vacuum, 
as discussed in \cite{Gandhi:2015xza}, the CC  probability has 
no sensitivity to $\dc$.}, which leads to degeneracies. Thus NC measurements, with their high statistics, 
are an important complementary tool to  probe CP and break degeneracies in new physics 
scenarios in conjunction with CC measurements.

\section{Neutral current measurements as a tool to break BSM physics degeneracies}

\begin{figure}[h]
\center
\includegraphics[width=0.49\textwidth]{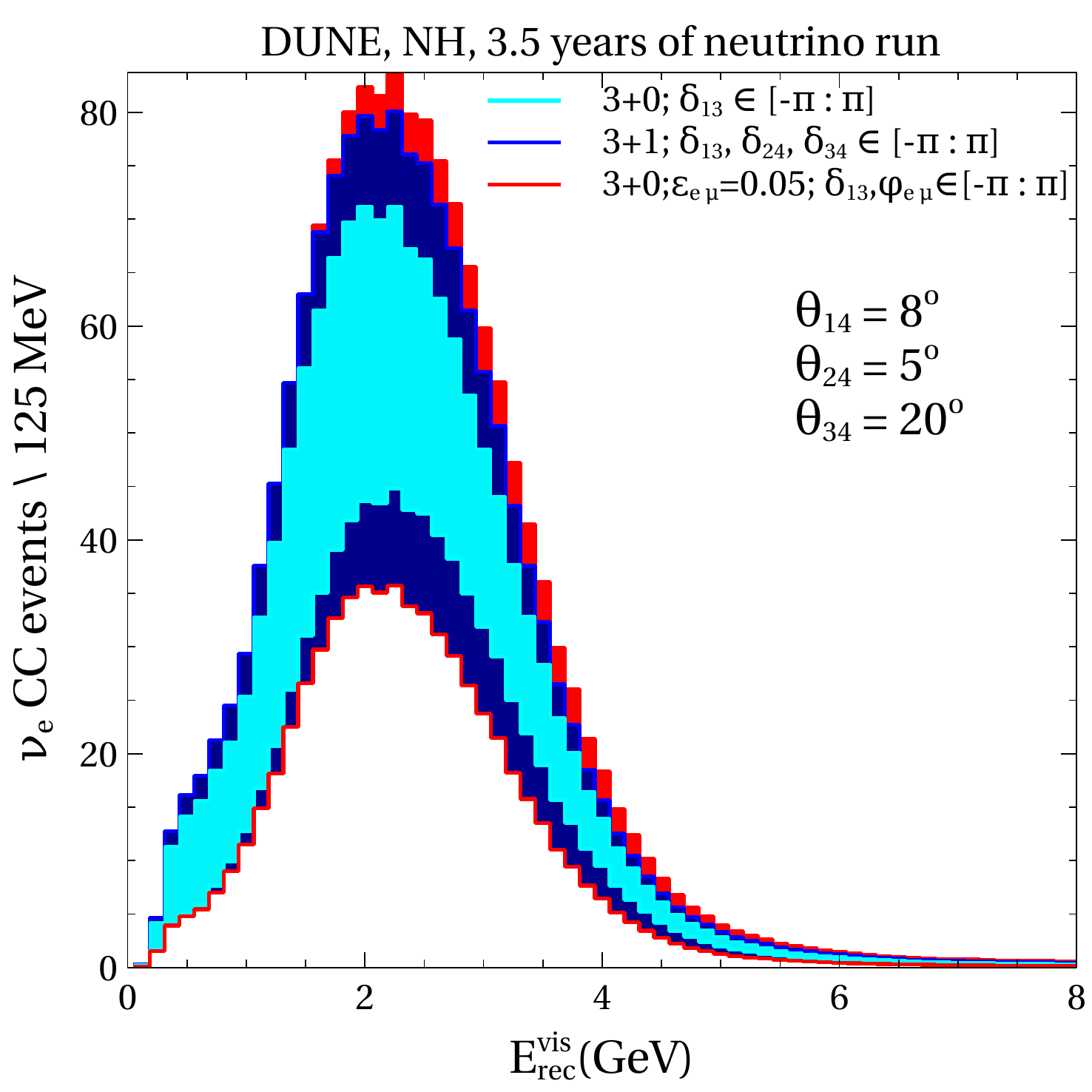}
\includegraphics[width=0.49\textwidth]{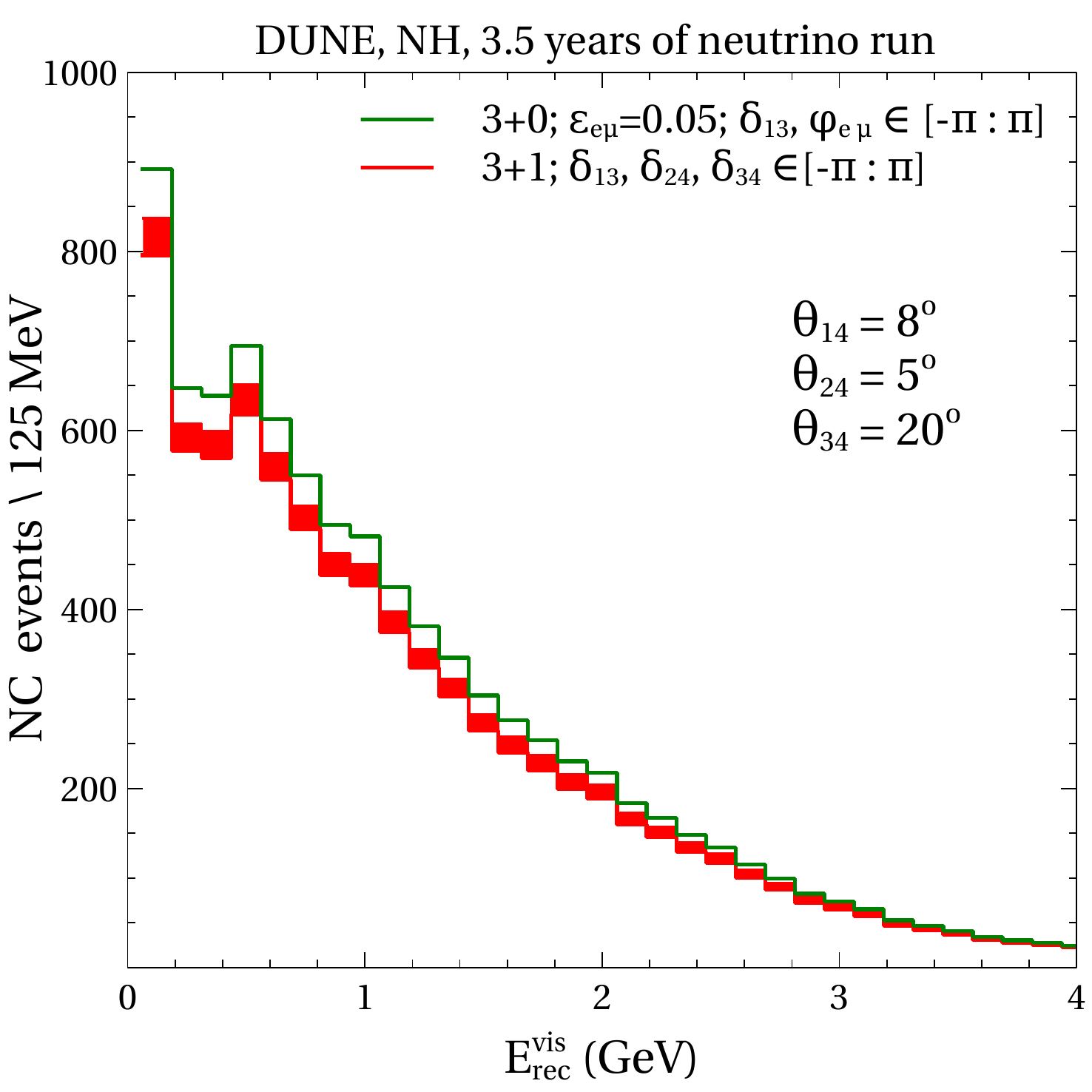}
\caption{\footnotesize{CC and NC events as a function of reconstructed 
neutrino energy in DUNE with 3.5 yrs of neutrino running.
The left panel corresponds to $\nu_{e}$ CC events 
for two different new physics scenarios, as well as for the standard 3+0 paradigm. 
The green line and the red band in the right panel show NC
neutrino events in the presence of propagation-related NSI, and in the presence 
of a sterile neutrino, respectively. 
In all cases the respective CP  phases have been varied over their full range 
of $[-180^\circ,+180^\circ]$. In the case of NSIs, 
$A(\delta_{e\alpha}\delta_{e\beta}+\epsilon_{\alpha\beta}
e^{i\phi_{\alpha\beta}})~(\alpha,\beta=e,\mu,\tau)$
represents the matter term in the effective Hamiltonian
in the presence of NSIs. Here, $A$ is the Wolfenstein matter term
and is given by $A(\rm{eV}^2)=0.76\times10^{-4}\rho(g/cc)E(GeV)$, 
$\rho$ being the matter density and $E$, the neutrino energy. 
The chosen example values of NSI and sterile 
parameters are shown in the key. The remaining NSI parameters are equal
to 0.}}
\label{fig-LE_CC_DUNEevents}
\end{figure}

In this section, we demonstrate the capability of NC events to break degeneracies 
which would otherwise arise in CC events, vis a vis new physics scenarios. While 
we choose propagation based  non-standard interactions (NSI) and a 3+1 sterile 
scenario to demonstrate our point, our conclusion will hold for any two new physics 
settings, one of which does not break 3+0 unitarity  (in this example, the propagation NSI) 
and another one which does (3+1 sterile). A similar conclusion would hold, 
for example, for NSI in propagation and neutrino decay, or NSI in propagation and NSI in production or detection (which inherently violate unitarity by adding to or depleting the source neutrino beam). 

Both  NSI arising during propagation  and extra sterile neutrino states 
affect $\nu_e$ CC events. From Fig. \ref{fig-LE_CC_DUNEevents} 
(left panel) we see that there is a wide range of possible spectra that can arise either 
from propagation NSI or an  extra sterile neutrino state (3+1 scenario). Shown also is 
the standard 3+0 scenario band.   NSI  affect the individual transition 
probabilities but the total oscillation probability of all the active flavours  
remains unity. On the other hand, in the presence of  extra sterile states, 
the total oscillation probability of the active flavours  becomes less than 
unity, leading to a depletion in NC  events in the presence of sterile  states 
compared to propagation NSI (the right panel of Fig. \ref{fig-LE_CC_DUNEevents}). 
Thus, NC events break the degeneracy seen in the CC event spectrum. 
We expect around 9345 NC total signal events in the case of 3+0 (or with propagation NSI present) 
in DUNE for 3.5 years of neutrino run. 
With 3+1 and sterile oscillation parameters corresponding to the right 
panel of Fig. \ref{fig-LE_CC_DUNEevents}, this number will 
deplete to $\sim(8306 - 8804)$ depending on the true values of the CP violating 
phases. Thus, a 6\% - 11\% reduction in the total NC signal event rate
is possible for $\tc\approx20^\circ$. We do quantitative analyses
in Sec. \ref{sensitivity}, to show that with a reduction in NC rates of this 
size, DUNE can distinguish between
the 3+0 (or propagation NSI) and the 3+1 scenarios at a 90\% C.L.

We note that as the sterile parameters become small, the 3+1 and 3+0 scenarios 
merge and become indistinguishable. In other words, the red band in the right panel
of Fig. \ref{fig-LE_CC_DUNEevents}  will tend to 
grow narrower and merge with the green solid line. 
Thus, the 3+1 parameters need to be such that a measurable difference 
in the NC rate can be attained.

\section{Constraints on the 3+1 paradigm: Sensitivity forecasts with DUNE}
\label{sensitivity}

In this section, we demonstrate the sensitivity of the DUNE experiment 
to exclude the 3+1 scenario using a combined  analysis of NC 
and CC measurements. We assume a 40 kt Liquid Argon 
detector and 3.5 years each of neutrino and anti-neutrino running. 
We have used the optimised beam profile, as described earlier in Section \ref{NCatLBL}. The CC and NC 
events due to such a beam have been shown in 
Figs. \ref{3+0-uncertainty} and \ref{fig-LE_CC_DUNEevents} .
We simulate data assuming that 3+0 is the true case i.e. we 
put the mixing parameters $\Delta m ^2_{41}, \ta, \tb, \tc, \db$ and $\dc$
equal to 0. Note that in such a situation $\da$ is $\dcp$.
We assume the hierarchy to be normal, $\tx=33.48^\circ$, 
$\ty=8.5^\circ$ and $\tz=45^\circ$. The mass-squared
differences $\Delta m^2_{21}$ and $|\Delta m^2_{31}|$ have been taken to be
$7.5\times10^{-5}\rm{eV}^2$ and $2.45\times10^{-3}\rm{eV}^2$ \cite{Fogli:2012ua, Gonzalez-Garcia:2015qrr, Forero:2014bxa}
respectively. The CP phase $\da$ is assumed to be $-90^\circ$, based on 
the recent hints from \cite{Abe:2017vif, Adamson:2016tbq}. We now 
fit this simulated data with events generated assuming the 3+1 scenario. 
We consider $\ta\in[0,12^\circ]$\footnote{The 90\% C.L. allowed range
for $\ta$ has been taken from \cite{Adamson:2016jku}.}, 
$\tc\in[0,50^\circ]$\footnote{Note that, the allowed range of $\tc$ from
\cite{Aartsen:2017bap} is $\tc\in[0,23^\circ]$ at 90\% C.L. However, in order to show 
the individual contributions from various channels we consider larger
values of $\tc$.},
$\tz\in[40^\circ,50^\circ]$, $\da~\rm{and}~\dc\in[-180^\circ,+180^\circ]$ 
and $\Delta m^2_{41}\in[0.1,10]~\rm{eV}^2$. Previously, we argued that
the results with NC data will not depend significantly on the parameters
$\tb$ and $\db$. However, the same is not true of the CC events i.e.
the $\nu_{\mu} \rightarrow\nu_e$ and $\nu_{\mu} \rightarrow \nu_\mu$
oscillation channels. Hence, in the fit, we vary $\tb\in[0,4^\circ]$\footnote{The results from 
IceCube \cite{TheIceCube:2016oqi} dictate the allowed range for $\tb$ at 90\% C.L.} 
and $\db\in[-180^\circ,+180^\circ]$. We assume the hierarchy to be known and hence do not 
consider the inverted hierarchy while fitting. We generate event spectra for various 
combinations of these 3+1 test oscillation parameters and then calculate the
binned Poissonian $\dxx$ between such test events spectra and the 
simulated 3+0 true events spectra (data). We have assumed $5\%$ normalisation 
error for the signal events and $10\%$ normalisation error for the background
events. The $\dxx$ are marginalised over these systematic uncertainties through 
the method of pulls.

\begin{figure}[h]
\center
\includegraphics[width=0.49\textwidth]{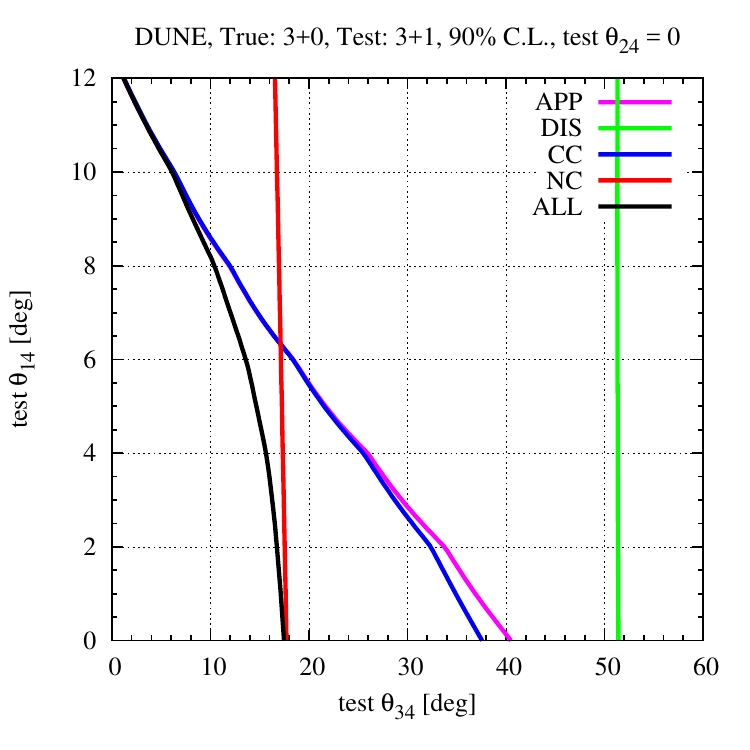}
\includegraphics[width=0.49\textwidth]{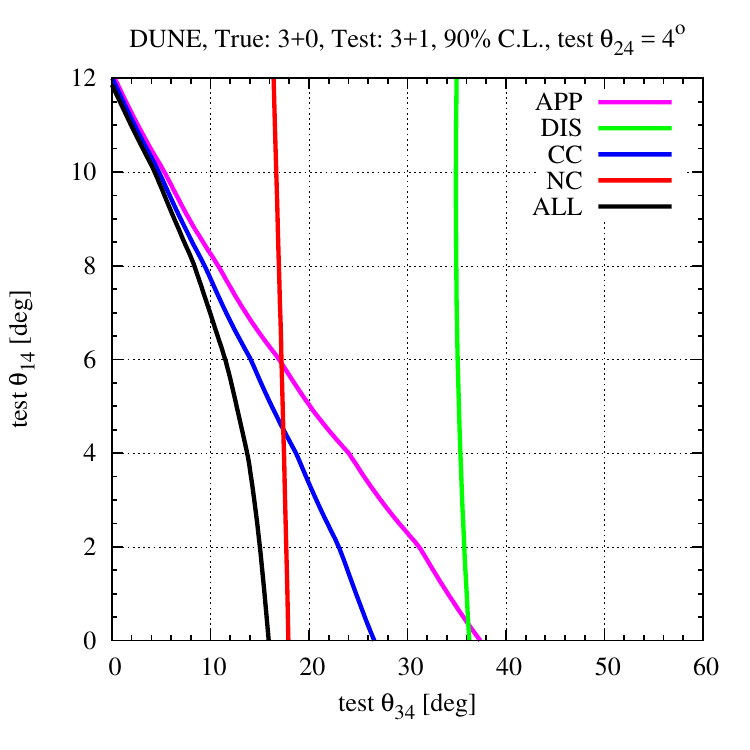}
\caption{\footnotesize{90\% ($\dxx=4.61$) C.L. contour plots in the test $\ta$ - test $\tc$ 
plane for different choices of test $\tb$. Left: test $\tb=0$ and
Right: test $\tb=4^\circ$. The true case has been
taken to be 3+0 and the test case is 3+1. The value of test $\Delta m^2_{41}$ is 
$1\rm{eV}^2$ for both the plots. The results are for the DUNE experiment
with 3.5 years each of neutrino and anti-neutrino running. For these figures, 
test ($\da,\db,\dc = -90^\circ,0,0$) i.e. the $\dxx$ has not been marginalised
over the test CP phases. We show 
results for the NC standalone data, appearance
standalone data, disappearance standalone data, appearance and disappearance 
data combined (CC) and finally, the CC and the NC data combined (``ALL").}}
\label{fig-sensitivityplots-1}
\end{figure}

In Fig. \ref{fig-sensitivityplots-1}, we show 
the sensitivity of the DUNE experiment to exclude the 3+1 
paradigm with NC and CC measurements. 
{\it We consider the test case of $\Delta m^2_{41} = 1~\rm{eV}^2$.} 
In producing these plots, we have not considered
the variation of the CP violating phases in the fit, so as to show
the effect of the mixing angles only. That is, we show the results 
corresponding to test $(\da, \db, \dc) = (-90^\circ,0,0)$. 
We show the $90\%$ C.L. limits (corresponding to $\dxx=4.61$ for a two-parameter fit) 
in the test $\ta$ - test $\tc$ plane for different values of test $\tb$. 
The left panel in Fig. \ref{fig-sensitivityplots-1} corresponds to the
choice of test $\tb=0$ and the right panel corresponds to test $\tb=4^\circ$, 
as depicted in the figure titles. We show results
for NC stand-alone data, appearance stand-alone data, disappearance
stand-alone data, appearance and disappearance combined ( i.e. CC
data) and finally all data i.e. CC and NC combined. This helps to
better understand the contribution that each type of data has in excluding  
the 3+1 scenario with respect to the given active-sterile mixing angle.
The regions that lie towards the increasing values of test $\ta$ and test $\tc$ are the 
ones for which DUNE can exclude 3+1 at $90\%$ C.L.  An examination of  the 
Fig. \ref{fig-sensitivityplots-1} allows us to draw some important conclusions:

\begin{itemize}

\item The NC data by itself constrains mainly the $\tc$ angle and this
constraint has a small dependence on the test values of the 
mixing angles $\ta$ and $\tb$. The most conservative exclusion of 
the $\tc$ angle corresponds to 
$\ta=0$ where $\tc\gtrapprox18^\circ$ is excluded by the data. The 
strongest bound of $\tc\gtrapprox16^\circ$ corresponds to $\ta=12^\circ$.

\item The appearance data are sensitive to all three active-sterile mixing angles. 
At $\tc=0$, $\ta\lessapprox12^\circ$ is allowed for both $\tb=0$ and $\tb=4^\circ$.
However, the constraints on $\tc$ are somewhat weak and strongly-correlated
with the values of test $\ta$. The weakest constraints are obtained for $\ta=0$,
which excludes values corresponding to $\tc\gtrapprox38^\circ$.

\item The disappearance data are mainly sensitive to $\tb$ and $\tc$. The constraints
are essentially independent of the value of test $\ta$. The strongest constraint, 
of $\tc\gtrapprox36^\circ$ being ruled-out, occurs when test $\tb=4^\circ$.

\item The combined NC+CC data are quite sensitive to $\tc$. If $\tb=4^\circ$
and $\ta\sim0$, then $\tc\gtrapprox16^\circ$ can be ruled out. For $\tb=4^\circ$
and $\ta\sim12^\circ$, DUNE data can rule out $\tc\gtrapprox0$.

\end{itemize}

It is quite evident from the above discussions that the NC data
have a marked advantage over the CC data in excluding 
the 3+1 paradigm when the mixing angles $\ta$ and $\tb$
are very small. If it so happens that the angles $\ta$
and $\tb$ are small but the angle $\tc$ is large, then, even though the
appearance and the disappearance data would not show any hints of 
new physics, there would be a clear evidence of new physics in the 
NC data.

\begin{figure}[h]
\center
\includegraphics[width=0.49\textwidth]{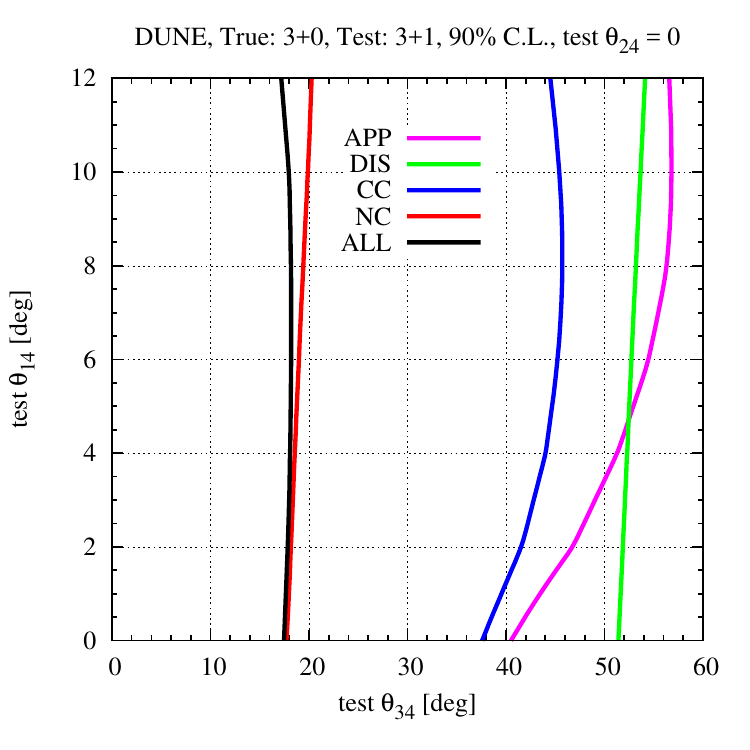}
\includegraphics[width=0.49\textwidth]{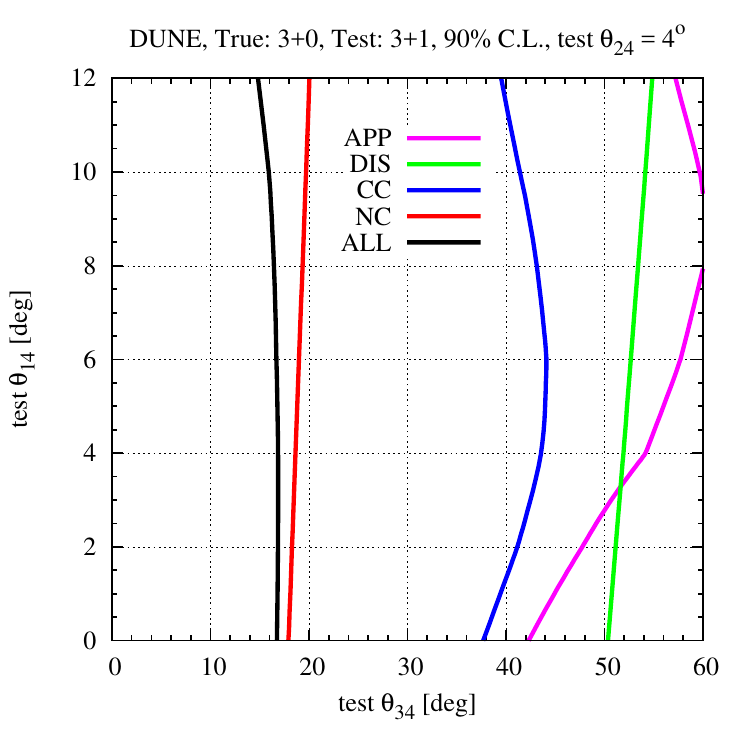}
\caption{\footnotesize{90\% ($\dxx=4.61$) C.L. contour plots in the test $\ta$ - test $\tc$ 
plane for different choices of test $\tb$. Left: test $\tb=0$ and
Right: test $\tb=4^\circ$. The true case has been
taken to be 3+0 and the test case is 3+1. The value of test $\Delta m^2_{41}$ is 
$1\rm{eV}^2$ for both the plots. The results are for the DUNE experiment
with 3.5 years each of neutrino and anti-neutrino running. For these figures, 
the $\dxx$ {\it has been} marginalised over the test CP phases. We show 
results for the NC standalone data, appearance
standalone data, disappearance standalone data, appearance and disappearance 
data combined (CC) and finally, the CC and the NC data combined (``ALL").}}
\label{fig-sensitivityplots-1a}
\end{figure}

In obtaining Fig. \ref{fig-sensitivityplots-1}, effects of the three CP phases were
not taken into account and each of the three of them were held fixed at 
their input true values. In Fig. \ref{fig-sensitivityplots-1a}, we repeat the same exercise as that in 
Fig. \ref{fig-sensitivityplots-1}, except that, for each test combination
of values of $\ta, \tb$ and $\tc$, we marginalise the $\dxx$ over the three
CP phases $\da, \db$ and $\dc$ and select the smallest $\dxx$. Thus, 
Fig. \ref{fig-sensitivityplots-1a} correctly takes into account the lack of 
knowledge regarding the CP violating phases. It can be seen that the results
due to CC appearance  are significantly 
affected because of marginalisation over the CP phases. This physics point was 
emphasised in \cite{Gandhi:2015xza, Dutta:2016glq}. While for the 
plots in Fig. \ref{fig-sensitivityplots-1}, a significant region of the given $\ta-\tc$ 
parameter space was ruled out by the appearance data; for the plots in Fig. \ref{fig-sensitivityplots-1a},
most of such $\ta-\tc$ region is allowed at 90\% C.L. This holds true especially at the 
larger values of $\ta$.
Thus, with CC data alone, DUNE cannot be expected to provide 
significant constraints on $\tc$. On the other hand, the effect of marginalisation
over CP phases on the NC data is small. Thus, NC data can 
decisively constrain the mixing angle $\tc$ even when the 
CP phases are unknown, as can be seen in the plots in Fig. \ref{fig-sensitivityplots-1a}.
Therefore, another advantage that the NC events have over  CC  is that they 
are more immune to the lack of knowledge regarding the CP phases.
{\it Even with CP violating phases present, it would be easier to rule out a moderately large
value of $\tc$ with the NC data compared to ruling out moderately
large values of $\ta$ and $\tb$ with the CC data.} Taking into account the marginalisation
over all the relevant mixing angles and the CP phases, the combined 
NC and CC data from DUNE can exclude the 3+1 paradigm for 
$\tc\gtrapprox18^\circ$. With reference to  Fig. \ref{fig-sensitivityplots-1} and 
Fig. \ref{fig-sensitivityplots-1a}, we note that 
in Fig. \ref{fig-sensitivityplots-1}, the most conservative estimate of 
$\tc$ corresponds to $\ta=0$. This is no longer true in Fig. \ref{fig-sensitivityplots-1a} 
where the most conservative constraints on $\tc$ occur at larger values
of $\ta$. This difference is stark in the case of CC data which reinforces the 
importance of CP phases in the CC channels. For NC too, this argument holds
true although the differences are much smaller in nature.

\begin{figure}[h]
\center
\includegraphics[width=0.49\textwidth]{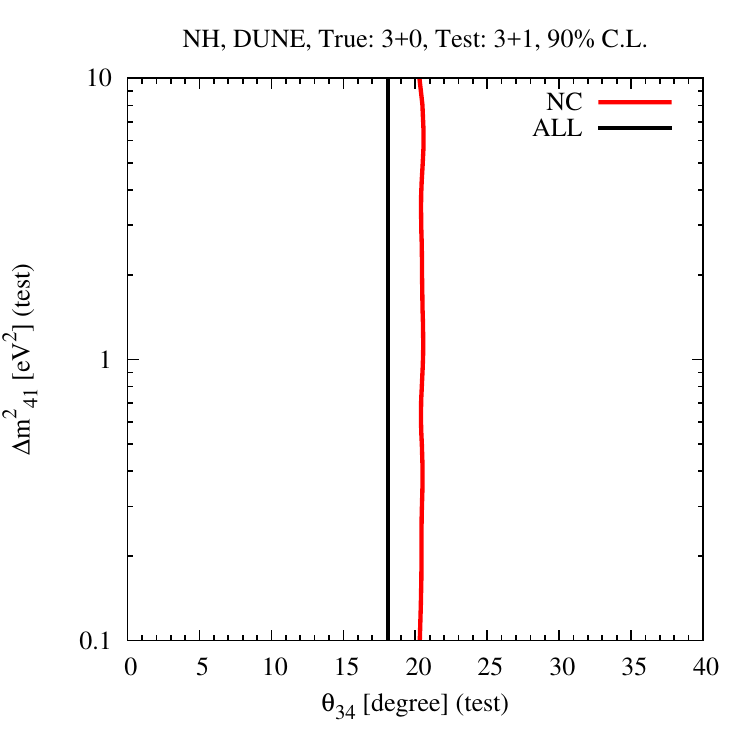}
\includegraphics[width=0.49\textwidth]{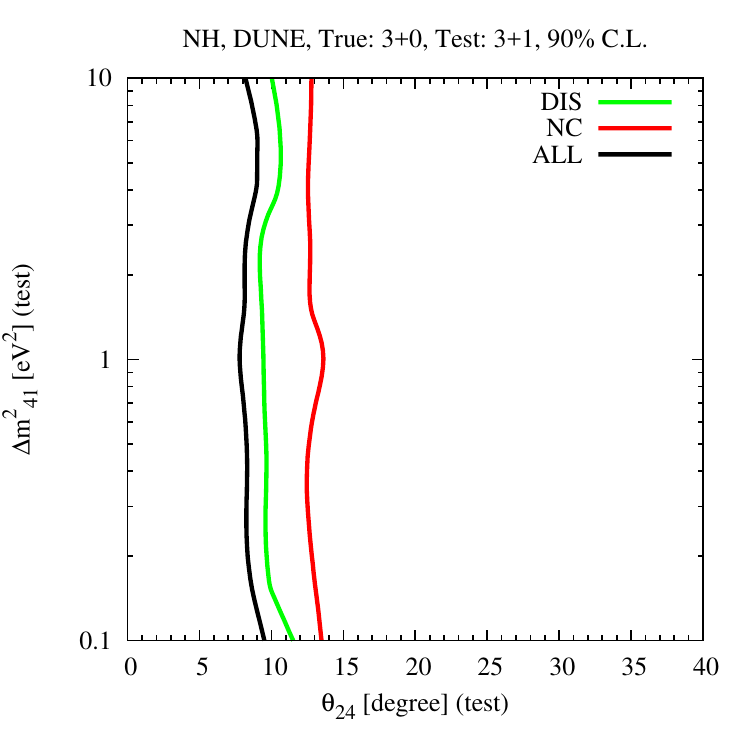}
\caption{\footnotesize{$90\%$ ($\dxx=4.61$) C.L. contour plots for 3+1 exclusion in the 
test $\Delta m^2_{41}$ - test $\tc$ space (left) and test 
$\Delta m^2_{41}$ - test $\tb$ space (right). The results are for the DUNE experiment
with 3.5 years each of neutrino and anti-neutrino running. The true case
has been taken to be 3+0 and the test to be 3+1. In the left panel, we show results
with the NC data, and NC and CC data combined (``ALL"). In the right panel,
we show results with the NC data, the disappearance data (``DIS") and the NC and CC data
combined (``ALL").}}
\label{fig-sensitivityplots-2}
\end{figure}

To show how the exclusion of the 
3+1 paradigm depends on the mass-squared difference 
$\Delta m^2_{41}$, we repeat the exercise done in Fig. \ref{fig-sensitivityplots-1} 
for test $\Delta m^2_{41}$ values ranging in $[0.1, 10]~\rm{eV}^2$.
We marginalise over the two mixing angles 
$\ta$ and $\tb$, in addition to the CP phases, and report the
minimum $\dxx$ as a function of test $\Delta m^2_{41}$ and 
test $\tc$. The results are shown in the left panel of Fig. \ref{fig-sensitivityplots-2}. 
Note that the other details regarding the simulation and assumptions 
on the oscillation parameters remain the same as those in Fig. \ref{fig-sensitivityplots-1}.
It is easy to see that the results do not depend much on the 
mass-squared difference $\Delta m^2_{41}$. At 90\% 
C.L., $\tc\gtrapprox18^\circ$ can 
be ruled out with the combined CC and NC data. With NC 
data alone, $\tc\gtrapprox20^\circ$ can be ruled out at 90\% C.L. 
NC data is most effective in constraining $\tc$. On combining the NC data 
with the CC data an improvement of $\approx2^\circ$ is seen. 

We show DUNE's ability to constrain the $\Delta m^2_{41}-\tb$ 
parameter space in the right panel of Fig. \ref{fig-sensitivityplots-2}.
We consider test $\Delta m^2_{41}$ values ranging in $[0.1, 10]~\rm{eV}^2$
and test $\tb$ values in $[0,40^\circ]$. In the fit, we marginalise over $\ta$
and $\tc$ and the three CP violating phases. It can be seen that most of the
sensitivity to the exclusion of $\tb$ comes from the disappearance data. With
the CC and NC data combined, DUNE can rule out $\tb\gtrapprox9^\circ\pm1^\circ$
depending on the test value of $\Delta m^2_{41}$. The current results from 
IceCube already exclude $\tb\gtrapprox4^\circ$ at 90\% C.L. for 
test $\Delta m^2_{41}\approx 0.5~\rm{eV}^2$ However, IceCube's $\tc$-constraint is
strongly correlated with the test value of $\Delta m^2_{41}$ and it can be
seen in \cite{TheIceCube:2016oqi} that for test $\Delta m^2_{41}\approx10~\rm{eV}^2$, 
the constraints from the IceCube data worsen to $\tb\gtrapprox45^\circ$ at 90\% C.L. 
DUNE, on the other hand, can provide a strong constraint on $\tb$ that is relatively 
independent of test $\Delta m^2_{41}$.

\section{Summary and concluding remarks}

This work attempts to examine how NC events  can synergistically aid the 
search for new physics and CP violation when combined with other measurements. 
We show that typically the NC events offer a window to CP phases and mixing 
angles that is complementary to that accessed by CC event measurements at 
both long and short baseline experiments.  They can break degeneracies existing 
in CC measurements, allowing one to distinguish between new physics that 
violates 3+0 unitarity and new physics that does not. NC events seem not to be 
affected greatly by matter effects which arise at energies and baselines relevant to 
DUNE, rendering analytical understanding  of new physics somewhat easier. They 
also aid in constraining parameters that are not easily accessible to CC measurements. 
Overall, in an experimental era when combined measurements can lead to significantly  increased
 precision and understanding, NC studies can play a valuable role in the search for new physics at neutrino detectors.

\section*{Acknowledgements}
We are grateful to Michel Sorel for providing us with the migration matrices used in this work, for a calculation probing a feature of the results obtained with these matrices, and for very helpful  discussions.
RG and BK thank Georgia Karagiorgi and Stephen Parke for very useful
conversations, and Mark Ross-Lonergan for some very helpful checks of
calculations of Long Baseline oscillation probabilities in 3+1.
RG acknowledges support in the form of a Neutrino Physics Center Fellowship
from the Neutrino Division and Theory Group at Fermilab. He, SP and SR also
acknowledge support from the XII Plan Neutrino Project of the Department of
Atomic Energy and the High Performance Cluster Facility at HRI.
SP thanks S\~ao Paulo Research Foundation (FAPESP)
for the support through Funding Grants No. 2014/19164-6 and
No. 2017/02361-1. He also thanks RG for the academic visit at Harish-Chandra
Research Institute during 2016 - 2017. This research was supported in part by the National Science Foundation under Grant No. NSF PHY11-25915. This manuscript has been authored by Fermi Research Alliance, LLC under Contract No. DE-AC02-07CH11359 with the U.S. Department of Energy, Office of Science, Office of High Energy Physics. The United States Government retains and the
publisher, by accepting the article for publication, acknowledges that the United States Government
retains a non-exclusive, paid-up, irrevocable, world-wide license to
publish or reproduce
the published form of this manuscript, or allow others to do so, for United
States Government purposes.

\appendix

\bibliographystyle{apsrev}
\bibliography{references.bib}

\end{document}